\documentclass[traditabstract]{aa}

\usepackage{graphicx}
\pdfoutput=1
\usepackage{amsmath}
\usepackage[varg]{txfonts}
\usepackage{natbib}
\usepackage[usenames,dvipsnames]{color}
\usepackage{enumerate}
\usepackage[normalem]{ulem} 
\usepackage{pifont}
\newcommand{\Mprim}{M_{1,\mathrm{i}}}
\newcommand{\Msec}{M_{2,\mathrm{i}}}
\newcommand{\Mdon}{M_{\mathrm{don}}}
\newcommand{\Macc}{M_{\mathrm{acc}}}

\newcommand{\Mpmz}{M_{\mathrm{PMZ}}}
\newcommand{\sepi}{a_{\mathrm{i}}}
\newcommand{\Pin}{P_{\mathrm{i}}}
\newcommand{\Pf}{P_{\mathrm{f}}}
\newcommand{\Porb}{P_{\mathrm{orb}}}
\newcommand{\Oorb}{\Omega_{\mathrm{orb}}}
\newcommand{\logg}{\log_{10} g}

\newcommand{\Tcond}{T_{\mathrm{cond}}}
\newcommand{\Kmin}{K_{\mathrm{min}}}
\newcommand{\aBHL}{\alpha_{\mathrm{BHL}}}
\newcommand{\aCE}{\alpha_{\mathrm{CE}}}
\newcommand{\lCE}{\lambda_{\mathrm{CE}}}
\newcommand{\lion}{\lambda_{\mathrm{ion}}}

\newcommand{\RRL}{R_{\mathrm{RL}}}
\newcommand{\Rd}{R_{\mathrm{d}}}

\newcommand{\qcrit}{Q_{\mathrm{crit}}}
\newcommand{\vw}{v_{\mathrm{w}}}
\newcommand{\vorb}{v_{\mathrm{orb}}}
\newcommand{\eiso}{\eta_{\mathrm{iso}}}

\newcommand{\ehydro}{\eta_{\mathrm{hydro}}}
\newcommand{\ebt}{\eta_{\mathrm{BT93}}}
\newcommand{\fbin}{f_{\mathrm{bin}}}
\newcommand{\flogP}{f_{\log P}}
\newcommand{\fC}{f_{\mathrm{C}}}

\newcommand{\Msun}{M_{\odot}}
\newcommand{\Rsun}{R_{\odot}}
\newcommand{\Lsun}{L_{\odot}}

\newcommand{\km}{\mathrm{km}}
\newcommand{\s}{\mathrm{s}}
\newcommand{\kms}{\mathrm{km\,s}^{-1}}
\newcommand{\days}{\mathrm{days}}

\newcommand{\Hy}{\mathrm{H}}
\newcommand{\C}{\mathrm{C}}

\newcommand{\Ox}{\mathrm{O}}

\newcommand{\Fe}{\mathrm{Fe}}

\newcommand{\Ba}{\mathrm{Ba}}

\newcommand{\Pb}{\mathrm{Pb}}

\newcommand{\X}{\mathrm{X}}
\newcommand{\Y}{\mathrm{Y}}
\newcommand{\ls}{\mathrm{ls}}
\newcommand{\hs}{\mathrm{hs}}
\begin{document}
   \title{Understanding the orbital periods of CEMP-s stars}

   \author{Carlo Abate
          \inst{1}
          \and
          Onno R. Pols\inst{2}
          \and
          Richard J. Stancliffe\inst{1}
          }

   \institute{Argelander-Institut f\"ur Astronomie der Universit\"at Bonn, Auf dem H\"ugel 71, D-53121 Bonn, Germany\\\email{cabate@uni-bonn.de}
   				\and
   				Department of Astrophysics/IMAPP, Radboud University Nijmegen, P.O. Box 9010, 6500 GL Nijmegen, The Netherlands
             }

   \date{Received ...; accepted ...}


\abstract
{
	{
	The chemical enrichments detected in carbon- and $s$-element-enhanced metal-poor (CEMP-$s$) stars are believed
	to be the consequence of a past episode of mass transfer from a now extinct asymptotic-giant-branch primary star.
	This hypothesis is borne out by the evidence that most CEMP-$s$ stars exhibit radial-velocity variations suggesting that they 
	belong to binary systems in which the companion is not directly visible.
 	}%
	{
	We use the orbital-period distribution of an unbiased sample of observed CEMP-$s$ stars to investigate the constraints it imposes
	on our models of binary evolution and on the properties of the metal-poor binary population in the Galactic halo.
	}%
	{
	We generate synthetic populations of metal-poor binary stars using different assumptions about the initial 
	period distribution and about the physics of the mass-transfer process, and we compare the predicted
	period distributions of our synthetic CEMP-$s$ stars with the observed one.
	}%
	{
	With a set of default assumptions often made in binary population-synthesis studies, the observed period 
	distribution cannot be reproduced. The percentage of observed CEMP-$s$ systems with periods shorter than
	about $2,\!000$ days is underestimated by almost a factor of three, and by about a factor of two between 
	$3,\!000$ and $10,\!000$ days. Conversely, about $40\%$ of the simulated systems have periods longer than 
	$10^4$ days, which is approximately the longest measured period among CEMP-$s$ stars. Variations in the
	assumed stability criterion for Roche-lobe overflow and the efficiency of wind mass transfer do not alter
	the period distribution enough to overcome this discrepancy.
	}%
	{
	To reconcile the results of the models with the orbital properties of observed CEMP-$s$ stars, one or 
	both of the following conditions are necessary: %
	($i$) the specific angular momentum carried away by the material that escapes the binary 
	system is approximately two to five times higher than currently predicted by analytical models and hydrodynamical
	simulations of wind mass transfer, %
	and ($ii$) the initial period distribution of very metal-poor binary 
	stars is significantly different from that observed in the solar vicinity and weighted towards periods
	shorter than about ten thousand days. Our simulations show that some, perhaps all, of the observed 
	CEMP-$s$ stars with apparently constant radial velocity could be undetected binaries with periods
	longer than $10^4$ days, but the same simulations also predict %
	that twenty to thirty percent of detectable binaries should have periods above this threshold, 
	much more than are currently observed. %
	}%
}
	\keywords{Binaries: mass transfer, stellar winds, angular-momentum loss, orbital periods. Stars: chemically peculiar, Population II. Galaxy: halo}

\nopagebreak
\maketitle

\section{Introduction}
\label{intro}

A significant proportion of the low-metallicity stars observed in the Galactic Halo are found 
to have abundances of carbon relative to iron more than ten times larger than in the Sun, that is%
\footnote{Given the elements X and Y and their number densities, 
$N_{\X,\Y}$, $[\X/\Y]=\log_{10}\left(N_{\X}/N_{\Y}\right)_{\star} - \log_{10}\left(N_{\X}/N_{\Y}\right)_{\odot}$, 
where $\star$ and $\odot$ indicate the abundances detected in the star and in the Sun, respectively.} %
$[\C/\Fe]>1.0$. These so-called carbon-enhanced metal-poor (CEMP) stars are a significant fraction of
the metal-poor population of the Halo and their proportion increases with decreasing metallicity,
making up more than $20\%$ of all metal-poor stars at $[\Fe/\Hy]<-3$ \cite[e.g.][]{Cohen2005,
Frebel2006, Lucatello2006, Carollo2012, Lee2013, Yong2013III, Placco2014}.
At $[\Fe/\Hy]>-3.5$ the majority of these carbon-rich stars also have strong enhancements in the element barium, 
which is predominantly produced by the \textit{slow} neutron-capture process. 
These objects are therefore classified as CEMP-$s$ stars. The main site of nucleosynthesis of $s$-elements
is the intershell region of thermally-pulsing asymptotic giant branch (TP-AGB) stars
\cite[e.g.][]{Gallino1998, Busso1999, Herwig2005, Romano2010, Prantzos2018}. However, the luminosities
and surface gravities of observed CEMP-$s$ stars prove that most of these objects, if not all, have not yet 
reached the AGB phase. On the other hand, the majority of CEMP-$s$ stars are found in binary systems 
\cite[][]{Aoki2007, HansenTT2016-2}, suggesting that the origin of the chemical enrichment 
is mass transfer from a binary companion that was once a TP-AGB star.

Insight into the nature of the mass-transfer mechanism can be gained from the study of the orbital properties of
these systems. However, information on orbital periods is hard to come by, particularly for long-period systems
for which an accurate orbital solution cannot be achieved without observations spanning many years. \cite{Lucatello2005-2}
were the first to suggest that all CEMP-$s$ stars are members of binary systems, based on a sample of 19 stars
with radial-velocity variations. They were able to calculate orbital solutions for ten of these systems. With
the exception of HE~$0024$--$2523$, which has a very short period of just 3.41 days, all
the systems have periods between a few hundred and a few thousand days. With this same data, and additional
radial-velocity data from many sources, \cite{Starkenburg2014} performed a maximum-likelihood analysis and concluded that
the binary fraction of the CEMP-$s$ stars is consistent with unity. They also placed a maximum period 
of around $10^4$ days on such systems, with an average period of around $500$ days. In addition, they showed that 
CH stars (the higher metallicity analogues of the CEMP-$s$ stars) have a similar, if not tighter, period distribution.

In their analysis of $13$ low-metallicity carbon stars, \cite{Jorissen2016} provided orbital solutions 
for an additional four CEMP-$s$ stars, finding periods of $400$--$3,\!000$ days in line with those
determined for previous systems. They also noted the similarity of the period range occupied by 
the CH and CEMP-$s$ stars, and in addition pointed out that the two
groups have a similar distribution in period-eccentricity space. Similar orbital properties 
and a binary fraction consistent with $100\%$ have been determined for barium stars,
a class of G and K barium-rich giants at solar metallicity, which are believed to form by
the same mass-transfer mechanism as CH and CEMP-$s$ stars \cite[e.g.][]{BoffinJorissen1988, 
McClure1990, Jorissen1998, VanderSwaelmen2017}.

\begin{figure}
		\includegraphics[width=0.49\textwidth]{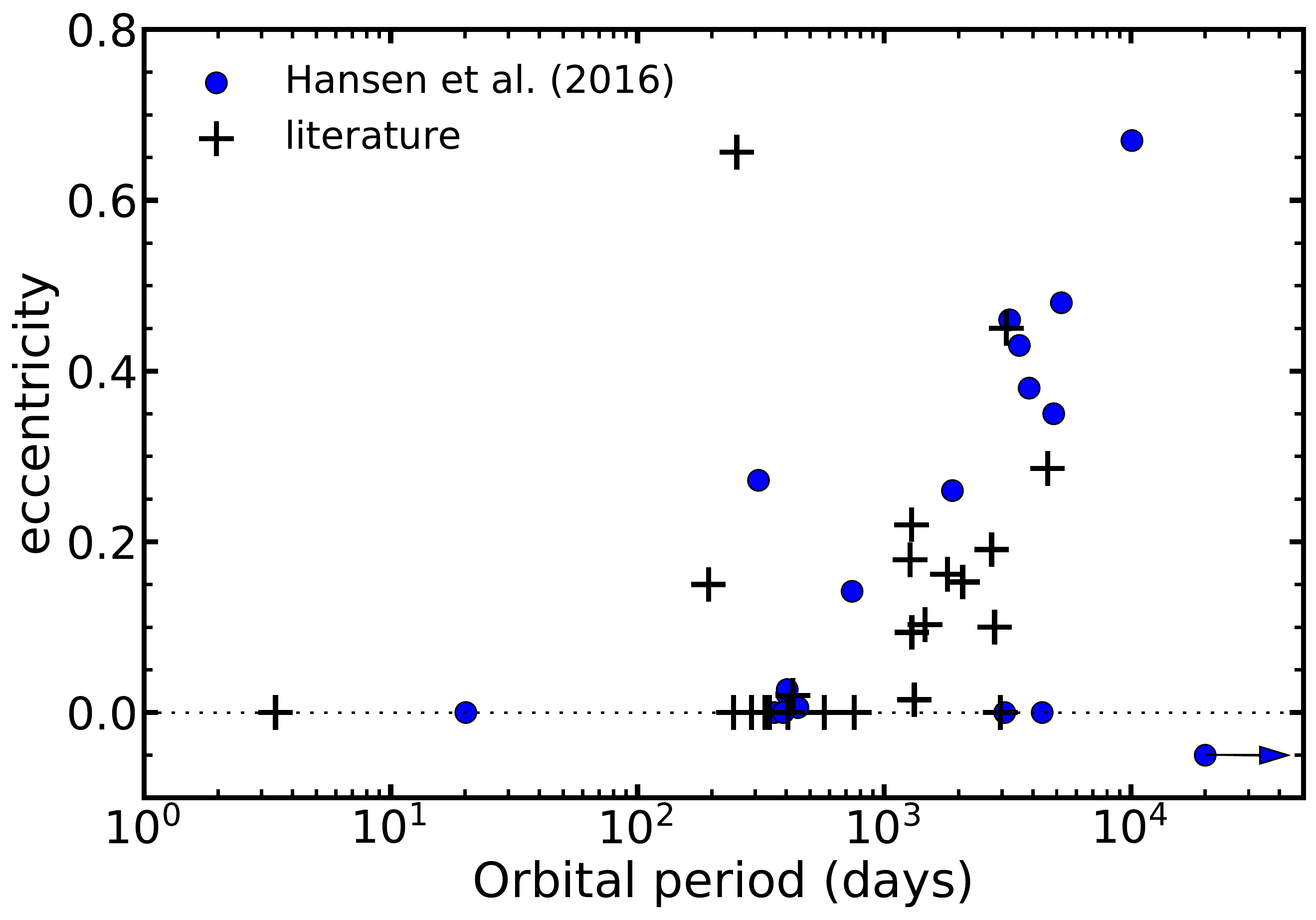}
		\caption{Period-eccentricity diagram of the binary systems detected by \citet[][blue-filled circles]{HansenTT2016-2} and of all other
		known binary CEMP-$s$ stars in the literature \cite[][black crosses]{Suda2008, Suda2011, Jorissen2016}.
				}
		\label{fig:fig5hansen}
	\end{figure}

\cite{HansenTT2016-2} built a sample of 22 CEMP-$s$ stars, selected only based on
their enhanced abundances of carbon and barium, and they monitored the radial velocities
of these systems monthly over a period of about $3,\!000$ days. They determined $17$ orbits independently
of other work, adding twelve CEMP-$s$ stars to the set of those with known orbital parameters.
Because this sample was chosen regardless of any previous detection of radial-velocity variations, 
it is not expected to be biased towards any period range and
it should be representative of the orbital properties of the overall CEMP-$s$ population. 
\cite{HansenTT2016-2} found periods between $20$ and $10,\!000$ days for $17$ out of $22$ systems 
(shown as blue-filled circles in Fig.~\ref{fig:fig5hansen}). Four stars are apparently single,
while one further system exhibits clear radial-velocity variations but it was impossible to 
determine its orbit (the point at negative eccentricity in Fig.~\ref{fig:fig5hansen}), 
indicating that the orbital period is probably very long.

Model predictions from binary population synthesis \cite[][]{Abate2015-3} show that the observed fraction of CEMP stars
among the very metal-poor stars of the SDSS/SEGUE survey can be reproduced at $[\Fe/\Hy]\lesssim-2.0$, but only with the contribution of binaries 
in a wide range of initial periods, up to a few times $10^5$ days. This also yields a wide distribution of current orbits, mostly 
in the range $10^3$ up to almost $10^6$ days \cite[][]{Izzard2009, Abate2013, Abate2015-3}. In particular,
\cite{Abate2015-3} demonstrated that the orbital-period distribution of the simulated binaries is
shifted towards periods longer by a factor of ten (on average) compared to the observed distribution.
Because the data available at the time was an inhomogeneous collection 
of orbital periods from the literature, the authors could not draw any definite
conclusions from this comparison. Either
the models should produce more CEMP stars in binary systems below a few thousand days, or, alternatively, 
the observed sample was biased towards short periods and most observed CEMP stars 
without an orbit determination should have periods longer than $10^4$ days.
The comparison with the unbiased sample of \cite{HansenTT2016-2}, which has periods in a range
consistent with previous results, has the potential to provide tighter
constraints on the simulations, in particular on the modelling of the mass-transfer process.

The nature of mass transfer in binary systems containing AGB stars is not well understood. 
Because mass transfer by Roche-lobe overflow (RLOF) from stars with deep convective envelopes is
in many cases unstable, it is likely an inefficient mechanism of mass transfer. Hence, systems 
with AGB donors have to be wide to avoid RLOF and the secondary stars accrete material from
the AGB winds. Unfortunately, our understanding of wind mass transfer is rather uncertain.
In situations where the wind speed is much faster than the orbital speed of the binary, 
the accretion can be described by the model of Bondi, Hoyle and Lyttelton \cite[][]{Hoyle1939, BoHo}, 
which results in low accretion efficiencies. However, AGB stars typically have slow winds
of not more than around $15\,\km\,\s^{-1}$ \citep{VW93}, making the situation much more complex.
Much effort has gone into modelling this type of mass transfer
\cite[e.g.][]{Shazrene2007, deValBorro2009, ChenZ2017, ChenZ2018,
deValBorro2017, Liu2017, Saladino2018-1}, which has led to the discovery of a new mode of mass transfer, dubbed
wind Roche-lobe overflow \cite[WRLOF,][]{Shazrene2007}. This mode relies on the fact that AGB winds
require the formation of dust to efficiently accelerate their winds. If the dust formation radius
lies sufficiently close to the Roche lobe, then the AGB wind moves very slowly and material can
be efficiently transferred to the companion through the inner Lagrangian point, with perhaps up
to half the ejected material being accreted \cite[][]{Shazrene2007, Abate2013}.

Coupled with the issue of mass transfer and mass loss from binary systems is the issue of 
angular-momentum loss and transfer. While the latter is most important for the subsequent 
evolution of the secondary, angular-momentum loss from the system as a whole will alter the 
binary orbit. In the Jeans approximation of a spherically symmetric wind, the ejected material
has the same specific angular momentum as the orbit of the wind-losing star and consequently 
the system widens in response to mass loss. However, in the case of a dense AGB outflow with 
velocity comparable to or lower than the orbital velocity of the binary the situation is 
certainly more complicated and both observations and hydrodynamical simulations show that 
the matter is not ejected isotropically (see e.g. \citealp{Karovska2005} 
for the observations of Mira AB, the prototypical detached binary system with an AGB donor star,
and \citealp{deValBorro2009}, \citealp{Shazrene2010}, \citealp{Liu2017} for the simulations). 
\cite{Jahanara2005} and, more recently, \cite{ChenZ2018} and \cite{Saladino2018-1} have 
performed hydrodynamical simulations to determine the amount of angular momentum carried
by the ejected material for different binary separations and mass ratios and a 
variety of assumptions about the input physics, such as the wind acceleration mechanism, chemical composition 
of the outflowing gas and its cooling efficiency. However, because of the complexity of the 
problem, a reliable model of how the amount of angular momentum lost by 
the binary system depends on all these parameters has yet to be developed.

In this work we use our binary population-synthesis model to investigate how different assumptions about
the accretion efficiency, angular-momentum loss, stability of Roche-lobe overflow and the initial
binary-parameter distributions modify the period distributions of synthetic CEMP-$s$ stars. 
In particular, we assume that the sample of CEMP-$s$ stars observed by \cite{HansenTT2016-2}
is representative of the overall population and we use the period distribution determined from this sample
as a case study to address the following questions:
    \begin{enumerate}
        \item What mechanism is responsible for mass transfer in binary systems and how efficient is it?
        \item How much angular momentum is carried away by the material that leaves the binary system?
        \item Is there a set of assumptions with which our binary-population-synthesis model reproduces the 
        orbital-period distribution observed in CEMP-$s$ stars?
    \end{enumerate}

\section{Models}
\label{model}

    In this study we use the binary population synthesis code $\texttt{binary\_c/nucsyn}$%
    \footnote{SVN revision r5045.}
    developed by \cite{Izzard2004, Izzard2006, Izzard2009, Izzard2018}.
	The starting point of our analysis is the default model set~A of \cite{Abate2015-3}, of which we describe the basic
	properties, assumptions, and selection criteria in Sect. \ref{sub:popsyn}. We introduce several 
    modifications in the input physics of this model~set and we investigate their consequences for the
    orbital-period distribution of the synthetic CEMP-$s$ population. 
    These modifications are discussed in Sections \ref{sub:RLOF}--\ref{sub:initial}.
    Table~\ref{tab:all_models} presents a list of all adopted model sets, together with
    the CEMP fractions they produce and a few numbers characterising their period distributions (which will be discussed in Sect. \ref{results}).

\subsection{Population-synthesis models}
\label{sub:popsyn}

    Following \cite{Abate2015-3}, in our simulated populations of binary stars the primary masses
    and separations ($\Mprim$ and $\sepi$) are logarithmically spaced over the intervals $[0.5, 8.0]\,\Msun$
    and $[5, 5\times10^6]\,\Rsun$, respectively, while the secondary masses, $\Msec$, are linearly
    distributed in $[0.1, 0.9]\,\Msun$.
    Our grid resolution is
    $N=N_{\mathrm{M}1} \times N_{\mathrm{M}2}\times N_{a}$, with $N_{\mathrm{M}1}=100$, $N_{\mathrm{M}2}=32$,
    $N_{a}=80$, giving a total grid of $256,\!000$ systems. In most of our simulations we consider
    circular orbits, although ten observed CEMP-$s$ stars have eccentricities greater than zero.
    The results of modelling eccentric systems are discussed in Sect.~\ref{sub:ecc}.
    The mass of the partial mixing zone, a parameter that determines the abundances of
    neutron-capture elements synthetised in the AGB phase \cite[][]{Karakas2010, Lugaro2012, Abate2015-1},
    is set to be equal to $\Mpmz=2\times10^{-3}\Msun$, which is the default of \cite{Karakas2010}. 
    This assumption has negligible effects on the total fraction of CEMP stars or the period distribution of
    synthetic CEMP stars. This is because the populations of simulated CEMP stars are
    dominated by binary systems with initial primary masses smaller than about $2\Msun$ 
    \cite[cf. Fig.~1 of][]{Abate2015-3}, and the amount of $s$-elements produced at these low masses
    does not depend much on the mass of the partial-mixing zone \cite[][]{Lugaro2012, Abate2015-3}.
    The wind velocity and mass-loss rate on the AGB are computed according to the empirical relations
    determined by \cite{VW93}. Wind velocities vary between $5$ and $15~\kms$,
    as in the paper of \cite{Abate2013}, except in our model set M7 in which we adopt a maximum of $7.5\,\kms$.

    The primary masses in our populations are initially distributed according to the solar-neighbourhood initial mass
    function (IMF) proposed by \cite{Kroupa1993}. In most of our models the distribution of initial mass ratios $q_{\mathrm{i}}=\Msec/\Mprim$ 
    is flat in the interval $[0,\,1]$, and the separation distribution is flat in $\log a_i$. 
    The total binary fraction, $\fbin$, is assumed to be unity in the considered range of $\sepi$, $[5, 5\times10^6]\,\Rsun$.
    Following the approach of \cite{Moe2017}, these assumptions translate into a constant
    binary fraction per decade of orbital period, d$f/$d$\log P\equiv \flogP=0.11$, because our
    separation distribution spans a range of $10^9$ days in orbital periods and the sum over all
    $\log\!P$-bins adds up to the total binary fraction, $\sum_P \flogP=\fbin$. Alternative separation distributions are
    explored in Sect. \ref{sub:initial}.

    As initial composition we adopt the solar abundance distribution of \cite{Asplund2009} scaled down to metallicity 
    $10^{-4}$ (that is $[\Fe/\Hy]\approx-2.2$). We assume that the transferred material is mixed throughout the accreting 
    star (hereinafter, the \emph{accretor}) to mimic the effect of non-convective mixing processes, such as
    thermohaline mixing, which is expected to be efficient in low-mass stars \cite[][]{Stancliffe2007, Stancliffe2008}.
    Our assumption will overestimate the dilution effect of these non-convective processes \cite[see e.g.][]{Matrozis2017-1}
    but it has a small impact on the final properties of the CEMP population, partly because most of our synthetic CEMP stars
    have undergone first dredge-up, which efficiently mixes the accreted material anyway \cite[][]{Abate2015-3}.
    
    We evolve our binary systems 
    with these initial conditions and we select the stars that after ten billion years have not yet become white dwarfs.
    We determine which of these stars are visible with a criterion based on their luminosity
    following the method described by \citet[Sect. 2.3]{Abate2015-3} with $V$-magnitude limits at $6$ and $16.5$.
    According to the selection criteria of \cite{HansenTT2016-2}, we flag a star as CEMP-$s$ when its carbon and 
    barium surface abundances are $[\C/\Fe]>1$ and $[\Ba/\Fe]>0.5$, respectively.

\begin{table*}[!t]
\caption{Input physics adopted in our calculated binary-population models and predicted fraction of CEMP stars.}
\label{tab:all_models}
\centering
\tiny
\begin{tabular}{ l l l l c  c  c c c c}
\hline
\hline
Model set&	$\qcrit$	&	Wind accretion	&	Angular-momentum		&	$\sepi$ distribution	& CEMP	  & \multicolumn{3}{c}{$\log_{10}\Pf$ at percentiles}& K-S\\
		&				&	mode			&	loss					&	         				& (\%)	          & 2.5\%  & 50\% & 97.5\% & $p$-value\\
\hline
M1	&	H02		&	WRLOF				&	isotropic wind				&	$\flogP=0.11$, ~~$\sepi/\Rsun\in[5,5\times10^6]$	& 5.4	          & 2.83   & 4.07 & 5.13 & 0.001 \\    
M2	&	CH08		&	WRLOF			&	isotropic wind				&	$\flogP=0.11$, ~~$\sepi/\Rsun\in[5,5\times10^6]$	& 6.0	          & 2.61   & 3.93 & 5.11 & 0.012 \\    
M3	&	$10^6$		&	WRLOF			&	isotropic wind				&	$\flogP=0.11$, ~~$\sepi/\Rsun\in[5,5\times10^6]$	& 6.7			  & 2.66   & 3.81 & 5.09 & 0.054 \\    
M4	&	H02			&	BHL  			&	isotropic wind				&	$\flogP=0.11$, ~~$\sepi/\Rsun\in[5,5\times10^6]$	& 4.0			  & 2.84   & 3.96 & 4.93 & 0.002 \\    
M5	&	CH08		&	BHL  			&	isotropic wind				&	$\flogP=0.11$, ~~$\sepi/\Rsun\in[5,5\times10^6]$	& 4.6			  & 2.57   & 3.84 & 4.87 & 0.026 \\    
M6	&	CH08		&	WRLOF			&	hydro						&	$\flogP=0.11$, ~~$\sepi/\Rsun\in[5,5\times10^6]$	& 5.8			  & 2.20   & 3.91 & 5.09 & 0.032 \\    
M7	&	CH08		&	WRLOF			&	hydro ($\vw=7.5\km\,s^{-1}$)&	$\flogP=0.11$, ~~$\sepi/\Rsun\in[5,5\times10^6]$	& 6.2			  & 1.92   & 3.87 & 5.39 & 0.044 \\    
M8	&	CH08		&	WRLOF			&	BT93						&	$\flogP=0.11$, ~~$\sepi/\Rsun\in[5,5\times10^6]$	& 5.5			  & 1.65   & 3.41 & 4.88 & 0.773 \\    
M9 	&	CH08		&	WRLOF			&	$\Delta J/J=2~(\Delta M/M)$	&	$\flogP=0.11$, ~~$\sepi/\Rsun\in[5,5\times10^6]$	& 6.8			  & 2.96   & 3.92 & 4.97 & 0.018 \\    
M10	&	CH08		&	BHL  			&	$\Delta J/J=2~(\Delta M/M)$	&	$\flogP=0.11$, ~~$\sepi/\Rsun\in[5,5\times10^6]$	& 5.1			  & 2.00   & 3.63 & 4.55 & 0.294 \\    
M11	&	CH08		&	WRLOF			&	$\Delta J/J=3~(\Delta M/M)$	&	$\flogP=0.11$, ~~$\sepi/\Rsun\in[5,5\times10^6]$	& 7.7			  & 1.92   & 3.87 & 4.92 & 0.033 \\   
M12	&	CH08		&	WRLOF			&	$\Delta J/J=6~(\Delta M/M)$	&	$\flogP=0.11$, ~~$\sepi/\Rsun\in[5,5\times10^6]$	& 10.1  		  & 1.75   & 3.64 & 4.81 & 0.276 \\   
M13	&	CH08		&	WRLOF			&	isotropic wind				&	\cite{Moe2017}										& 5.0		  	  & 2.62   & 3.99 & 5.18 & 0.006 \\    
M14	&	CH08		&	WRLOF			&	isotropic wind				&	$\flogP=0.15$, ~~$\Pin/\days\in[10,2\times10^4]$	& 6.5			  & 2.54   & 3.67 & 4.44 & 0.108 \\
M15	&	CH08		&	BHL, $\aBHL=10$	&	isotropic wind				&	$\flogP=0.11$, ~~$\sepi/\Rsun\in[5,5\times10^6]$	& 7.2			  & 2.25   & 4.26 & 5.54 & 0.001 \\   
M16	&	CH08		&	BHL, $\aBHL=10$	&	$\Delta J/J=2~(\Delta M/M)$	&	$\flogP=0.11$, ~~$\sepi/\Rsun\in[5,5\times10^6]$	& 7.5			  & 2.01   & 4.06 & 5.36 & 0.005 \\   

\hline
\end{tabular}
\tablefoot{
			H02: as defined by \cite{Hurley2002}. CH08: table based on the results of \cite{Chen2008}. %
			Hydro: fit to hydrodynamical simulations from the literature \cite[][]{Jahanara2005,ChenZ2018,Saladino2018-1} as in Eq. (\ref{eq:hydro}). %
			BT93: fit to the ballistic simulations of \cite{Brookshaw1993} as in Eq. (\ref{eq:BT93}). %
			With the exception of M7, all sets assume terminal AGB wind velocity $\vw=15\km\,\s^{-1}$. %
			All models using the BHL approximation assume $\aBHL=1.5$, except M15 and M16.\\
}
\end{table*}

\subsection{Stability of Roche-lobe overflow}    
\label{sub:RLOF} 
   
    Roche-lobe overflow from AGB donors is believed to be generally unstable, except in some cases
    when the donor is less massive than its companion or the mass of the convective envelope is small 
    compared to that of the core \cite[][]{Hjellming1987}. This is because AGB stars expand in response 
    to mass loss because of their large convective envelopes \cite[][]{Paczynski1965}, whereas the
    Roche-lobe radius shrinks in response to mass transfer when the donor is more massive than its companion
    \cite[][]{Paczynski1965, Paczynski1971}.
    Favourable conditions for stable RLOF are rarely met in the formation process of CEMP-$s$ stars.
    Low-metallcitity AGB stars of initial masses above $0.9\,\Msun$ efficiently dredge up nuclear-processed 
    material to the surface \cite[][]{Stancliffe2009, Karakas2010, Lugaro2012}. By contrast, their binary
    companions need to be low in mass, $\Msec\leq 0.85\,\Msun$ \cite[][]{Abate2015-3}, otherwise after 
    accreting a few hundredths of a solar mass of material they rapidly evolve and become white dwarfs 
    before $10$--$12$ Gyr, which is approximately the age of the Galactic-halo population. In that 
    case they would not be observable today as CEMP-$s$ stars. Consequently, the mass ratio 
    in most potential progenitors of CEMP-$s$ stars is $M_2/M_1<1$ during the whole
    evolution and therefore the RLOF is in most cases unstable. The binary system is then
    believed to evolve into a common-envelope phase \cite[][]{Paczynski1976} during which there 
    is no significant accretion of material on to the companion star \cite[][]{RickerTaam08}.
    
    Our binary population-synthesis model determines whether RLOF is stable by comparing the mass ratio between
    the donor and the accretor, $Q=\Mdon/\Macc$, with a critical value, $\qcrit$: if $Q>\qcrit$ during RLOF,
    the systems undergo common-envelope evolution. When the donor star is a giant, $\qcrit$ is calculated with
    Eq. 57 of \citet[][Sect. 2.6.1]{Hurley2002} and scales with the fifth power of the ratio between
    the core and total masses of the donor. %
    When the RLOF process is unstable, we model the common-envelope evolution according to
    equations 69--78 of \citet[][Sect. 2.7]{Hurley2002}, in which we assume $\aCE=1.0$, $\lion=0.0$,
    and $\lCE$ is computed as in Eq. A.1 of \citet[][Appendix~A]{Claeys2014}.

    In recent years, the response of giant stars to
    mass loss has been investigated by several authors who have argued that stars with convective 
    envelopes may expand less, and less rapidly, than previously derived with simplified models
    \cite[e.g.][]{Chen2008, Ge2010, Woods2011, Passy2012-1, Passy2012-2, Ge2015, Pavlovskii2015}.
    \citet[][tables 1 and 2]{Chen2008} provide critical mass ratios for stable RLOF for different 
    primary masses, stellar radii, and mass accretion efficiencies. These values are generally 
    higher than the $\qcrit$ implemented in our code, which implies that RLOF from AGB donors 
    may be more stable than, for example, in the simulations of \cite{Abate2015-3}.
    Therefore, we use the results of \cite{Chen2008} to construct a table (R.G.~Izzard, priv. comm.) 
    which can be interpolated by our population-synthesis code to determine the stability of RLOF 
    in binary systems with an alternative criterion to that of \cite{Hurley2002}. Hereinafter we 
    will refer to the former as the ``CH08 criterion'' of RLOF stability and to the latter as the 
    ``H02 criterion''. The CH08 criterion is adopted in most of our model sets (see Table \ref{tab:all_models}).

    As we discuss in Sect. \ref{res:RLOF}, models with more stable RLOF from AGB donors predict a 
    larger number of CEMP-$s$ stars with orbital periods between a few hundred and a few thousand days. 
    To test the maximum possible effect of increased RLOF stability on the period distribution of 
    synthetic CEMP-$s$ stars, in our model set M3 we impose that RLOF from AGB stars is always 
    stable by setting an arbitrarily high $\qcrit$ (namely, $\qcrit=10^6$).

\subsection{Accretion efficiency of wind mass transfer}
\label{sub:accretion}    

    Because of the above-mentioned constraints on the stability of the RLOF process, mass transfer from AGB donors
    and the consequent formation of CEMP-$s$ stars is generally considered to occur by accretion of stellar winds.
    The efficiency of this process as a function of the masses of the two stars and of the orbital separation
    is not well understood.
    Population-synthesis studies often use the Bondi-Hoyle-Lyttleton model \cite[][]{Hoyle1939, BoHo, Bondi1952, Edgar2004}
    to determine the accretion efficiency of wind mass transfer. This prescription is adopted in our model sets labelled with
    BHL in Table~\ref{tab:all_models}, which compute the wind accretion rate using equation 6 of \cite{BoffinJorissen1988}:
    \begin{equation}
        \beta_{\mathrm{BHL}} = \frac{\aBHL}{2\sqrt{1-e^2}} \cdot 
        						\left(\frac{GM_\mathrm{acc}}{a~v_\mathrm{w}^2}\right)^2~\left[1 + \left(\frac{v_{\mathrm{orb}}}{v_{\mathrm{w}}}\right)^2\right]^{-\frac{3}{2}}
        ~,~\label{eq:BHL}
    \end{equation} 
    where $G$ is the gravitational constant, $v_\mathrm{w}$ and $v_\mathrm{orb}$ are the wind and orbital velocities, respectively,
    $e$ is the eccentricity, and $\aBHL$ is a numerical constant between $1$ and $2$ (by default it is equal to $1.5$ in our models).
    
    The BHL model is appropriate under the assumption that the orbital velocity of the accretor is much smaller than the wind velocity.
    AGB winds do not usually fulfil this condition. Their detected velocities vary between approximately $3$ and $30\,\km\,\s^{-1}$
    \cite[e.g.][]{VW93, vanLoon2005, Goldman2017}, which are comparable to the orbital velocities of binary stars 
    of total mass in the range $1-3\,\Msun$ and periods up to about $30,\!000$ days. In wider systems AGB winds are
    typically faster than the orbital velocities of the donor stars.
    
    \cite{Abate2013} use the results of detailed hydrodynamical calculations \cite[][]{Shazrene2007, Shazrene2012, Shazrene2010}
    to develop a simplified model of wind Roche-lobe overflow (WRLOF), a mode of mass transfer
    in which it is the slow and dense wind of the donor star, rather than the star itself, that fills the Roche lobe and 
    is transferred on to the binary companion. We refer to \cite{Abate2013} for a complete description of their WRLOF model
    and here we summarise the basics. 
    AGB winds are attributed to a combination of stellar pulsations,
    which create the conditions for dust condensation at some distance $\Rd$ from the surface of the AGB star, 
    and radiation pressure on dust grains. These are accelerated beyond the escape velocity and, by dynamical
    collisions with the gas, transfer a net outward momentum to the wind particles
    \cite[e.g.][]{Freytag2008, Nowotny2010, Bladh2012, Bladh2015, Hofner2015}.
    If $\Rd$ is greater than or comparable to the Roche-lobe radius $\RRL$ of 
    the donor, the AGB wind is slow inside the Roche lobe and is gravitationally focused through the inner
    lagrangian point $L_1$ and transferred with high efficiency to the secondary star. 
    The dust-formation radius, $\Rd$, is a function of the effective temperature of the star and of the
    condensation temperature of the dust \cite[][]{Hofner2007}. The latter depends on the chemical composition
    of the dust and it is assumed to be $1500\,$K and $1000$~K for carbon- and oxygen-rich dust, respectively
    \cite[][]{Hofner2009}. Because the dust composition of low-metallicity AGB stars is complex and often
    very uncertain \cite[e.g.][]{Boyer2015III, Boyer2015II}, in our model the condensation temperature,
    $\Tcond$, is treated as a free parameter. 

    In their hydrodynamical calculations of WRLOF, \cite{Shazrene2007} originally adopt $\Tcond=1000\,$K,
    which is also the value used by \cite{Abate2015-3} to maximise the fraction of CEMP stars in 
    their synthetic populations of metal-poor stars. 
    Indeed, a low dust-condensation temperature implies that the range in which WRLOF takes place is 
    shifted towards longer separations compared to the case of a higher $\Tcond$, because dust forms 
    further away from the star and hence $\Rd$ is larger. Consequently, the number of binary systems
    that undergo efficient wind mass transfer at long separations increases and so does the CEMP fraction%
    \footnote{We refer to Sect.~5 of \citealp{Abate2013} for a discussion about the consequences of varying $\Tcond$.}.
    However, a large proportion of these systems are formed at much longer periods than observed \cite[][]{Abate2015-3}.
    Following \cite{Abate2013}, here we choose to assume $\Tcond=1500$~K because after about five
    thermal pulses our model AGB stars have $\C/\Ox>1$ at the surface and hence the dust formed in their outflow
    is also likely carbon rich.

\subsection{Angular-momentum-loss model}    
\label{sub:AM}    
    
    The variation in orbital angular momentum caused by mass loss in a binary system can be parameterised as
    \begin{equation}
        \dot{J} = \eta \left( \dot{M}_{\mathrm{don}}-\dot{M}_{\mathrm{acc}} \right) a^2 \Oorb
        ~,~\label{eq:jdot}
    \end{equation} 
    where $\Oorb$ and $a$ are the orbital angular velocity and separation of the binary, respectively,
    $\dot{M}_{\mathrm{don}}$ and $\dot{M}_{\mathrm{acc}}$ are the mass-loss and mass-accretion rates of the donor
    and the accretor, respectively, hence their difference is the total mass lost by the system per unit time,
    and $\eta$ is a parameter identifying the specific angular momentum carried away by the expelled material per unit mass.
    
    In our model sets M1--M5, M13 and M14, the variations of orbital angular momentum because of wind mass loss are
    computed according to the Jeans approximation of an isotropic, spherically-symmetric wind, as e.g. in
    Eq.~(4) of \cite{Abate2013}. This is a valid approximation in the case of fast winds, with velocities much 
    larger than the orbital velocity of the binary. In this approach the specific angular momentum of the ejected 
    material is
    \begin{equation}
    	\eiso = \frac{1}{\left(1+Q\right)^2}
    	~.~\label{eq:jeans}
    \end{equation}
	This mode of mass loss always results in expansion of the orbit.
    
    In contrast with the isotropic-wind approximation, a variety of observations 
    \cite[e.g.][]{Karovska1997, Castro-Carrizo2002, Karovska2005, Karovska2010} and hydrodynamical simulations 
	\cite[e.g.][]{Theuns1993, Nagae2004, Shazrene2007, ChenZ2017, deValBorro2017, Liu2017} show that
    the geometry of the wind lost in binary systems with AGB donor stars is in many cases not spherical,
    but focussed into the orbital plane. 
    As a consequence, the angular momentum carried away by the ejected material may be larger than
    predicted by the Jeans mode.
    This in turn can cause the binary orbit to shrink rather than expand \cite[][]{ChenZ2018,Saladino2018-1}.
    Taking into account such enhanced angular-momentum loss may therefore help to explain the short orbital periods
    of CEMP-$s$ binaries and related systems.
    For example, \cite{Izzard2010} find that the observed
    period-eccentricity distribution of barium stars, which are often considered the solar-metallicity 
    analogs of CEMP-$s$ stars, can be reproduced if the ejected material carries
	away at least two times the average specific orbital angular momentum of the binary system.
    \cite{Abate2015-1} reach a similar conclusion in their effort to simultaneously model 
    the chemical composition and the orbital period of $15$ observed CEMP-$s$ stars. 
    
    By including the formalism of \citet[][Eq.~2]{Izzard2010} into our Eq.~(\ref{eq:jdot}), we obtain the following
	\begin{equation}
		\dot{J} = \gamma \times \frac{Q}{(1+Q)^2} \left( \dot{M}_{\mathrm{don}}-\dot{M}_{\mathrm{acc}} \right) a^2 \Oorb~~,
		\label{eq:gamma}
	\end{equation}
	from which follows the relation $\eta = \gamma \, Q \, (1+Q)^{-2}$.
    \cite{Izzard2010} and \cite{Abate2015-1} adopt $\gamma=2$. In this work,
    we test different values of the constant $\gamma$, namely $\gamma=2,3,$ and $6$, to qualitatively
    investigate how strong the angular-momentum loss has to be in order to reproduce the 
    period distribution of CEMP-$s$ stars%
	\footnote{For comparison, in this formalism $\gamma=Q^{-1}$ for an isotropic wind.} .%
    
	The choice of a constant $\gamma$ in Eq. (\ref{eq:gamma}) implies that the specific angular momentum 
	of the ejected material does not depend on the orbital period of the system. Hence, a binary system
	so wide that the gravitational influence of the companion star on the wind of the donor is negligible 
	loses the same amount of angular momentum as a close binary, if the total ejected mass in the two cases is the same. 
	A step towards a more physical description of the process is to include a dependence
	of $\eta$ on the orbital properties of the binary system. 
	For this purpose we use the results of the hydrodynamical simulations of \cite{Jahanara2005}, \cite{ChenZ2018} and 
	\cite{Saladino2018-1}, in which the angular-momentum loss rates from binaries interacting via stellar winds are 
	computed explicitly. \cite{ChenZ2018} and \cite{Saladino2018-1} present simulations of low-mass binaries 
	interacting via the winds of their AGB donor stars. Despite the different assumptions made in these studies, 
	the specific orbital angular momentum of matter lost from the system in both studies is very similar when it is 
	expressed as a function of the ratio of the terminal wind velocity and the orbital velocity, $\vw/\vorb$ 
	\cite[see][]{Saladino2018-1}. 
	\cite{Jahanara2005} present more generic simulations of wind mass transfer, of which those labelled as 
	`radiatively driven' are the most applicable to AGB winds. They also find that the specific angular momentum 
	of the ejected matter depends on the ratio of the wind velocity and the orbital velocity. 

	\cite{Saladino2018-1} find that the results of all three sets of simulations can be represented fairly well by a 
	simple relation,
	%
    \begin{equation}
		\ehydro = \eiso + \frac{1.2 - \eiso}{1 + (2.2 \vw/\vorb)^3}~~,
		\label{eq:hydro}
    \end{equation}
	%
	where $\vw$ and $\vorb$ are the wind and orbital velocities, respectively, and $\eiso$ is the 
	specific angular momentum for isotropic mass loss given by Eq.~(\ref{eq:jeans}). The second term in Eq.~(\ref{eq:hydro}) 
	gives $\ehydro$ a dependence on the separation through $\vorb$. 
    In binary systems with very long orbital periods the ratio $\vw/\vorb$ is large, consequently the first
    term of Eq. (\ref{eq:hydro}) dominates and the angular momentum lost by the system
    is the same as in the isotropic-wind model. By contrast, for shorter orbital separations the ratio 
    $\vw/\vorb$ decreases (because $\vw$ is constant while $\vorb$ increases) and consequently the contribution 
    of the second term in Eq. (\ref{eq:hydro}) becomes stronger.
    
    \begin{figure}[!t]
        \includegraphics[width=0.5\textwidth]{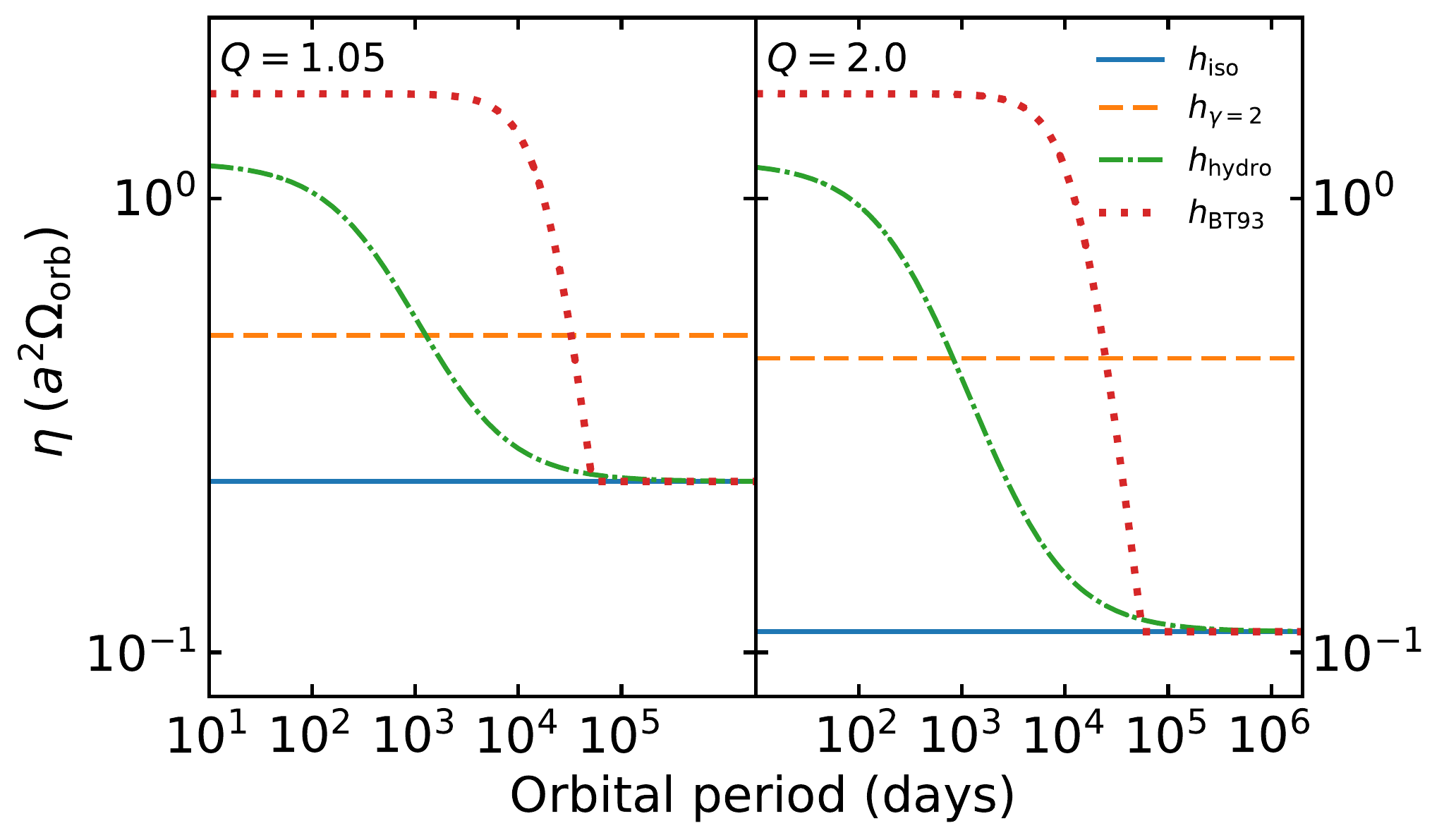}
	    \caption{Specific angular momentum, $\eta$ (in units of $a^2\Omega_{\rm{orb}}$), as a function of the orbital period of 
	   			binary systems with fixed primary mass, $\Mdon=0.9\Msun$, and mass ratios $Q=1.05, 2$ (left
	   			 and right panels, respectively). The solid, dashed, dot-dashed and dotted lines show,
	   			respectively, the profiles of $\eta$ as computed with Eqs. (2), (3) with $\gamma=2$,
	   			(4) and (5).}
    \label{fig:eta-vs-P}
    \end{figure}

    Alternatively, we can use the results of \cite{Brookshaw1993} who studied the angular-momentum loss from 
    binary systems in which the wind is modelled by ballistic calculations of
    test particles. These calculations ignore the effects caused by gas pressure and radiative acceleration. 
    This is permissible for fast winds, but it is a poor representation of slow and dense AGB winds. Because both
    these phenomena will tend to make the outflow more isotropic, the results of this ballistic study can be taken to
	give an upper limit to the amount of angular momentum lost by a stellar wind. We fit the results of
	\cite{Brookshaw1993} as a function of the mass ratio and $\vw/\vorb$ (see Appendix~\ref{appendixA}), 
	similarly to Eq. (\ref{eq:hydro}):
	%
    \begin{equation}
        \ebt = \mathrm{max}\left\{
        				~\eiso~, ~
        				\cfrac{1.7}{1 + [(0.6 + 0.02 Q) \cdot \vw/\vorb]^6}~
        				\right\}~~.
        \label{eq:BT93}
    \end{equation}
	%
	With Eq. (\ref{eq:BT93}) wide binary systems evolve as for an isotropic outflow,
	whereas in closer binaries the angular momentum lost is significantly higher. The
	transition between these two regimes is considerably steeper than in Eq.~(\ref{eq:hydro}), as shown by
	Fig.~\ref{fig:eta-vs-P}, in which we plot the specific angular momentum of the ejected material
	for different models and mass ratios ($Q=1.05$ and $2$ in the left and right panels, respectively)
	as a function of the orbital period in binary systems with primary mass $\Mdon=0.9\Msun$.
	Figure \ref{fig:eta-vs-P} also shows that in the models with constant $\gamma$ and with an isotopic wind, 
	$\eta$ does not depend on the orbital period and consequently the variations of angular
	momentum are only determined by the total mass that is lost by the system.
	
	The angular-momentum loss rates given by Eqs.~(\ref{eq:jeans})--(\ref{eq:BT93}) represent only the 
	angular-momentum loss from the \emph{orbit}, and do not include a possible contribution from the loss of rotational 
	angular momentum in the wind of the AGB donor star. The latter is accounted for separately in our code.
	In case the spin of the mass-losing star is tidally locked to the orbit, this rotational angular-momentum loss 
	is effectively also taken out of the orbit. 
	This is important in binaries that are close enough for tidal friction to occur on a timescale shorter 
	than the mass-loss timescale, and can result in additional orbital shrinkage. 
	In our binary population synthesis code, tidal friction and angular momentum transfer between the 
	stars and the orbit is calculated explicitly \cite[][]{Hurley2002} and the effect of spin angular-momentum loss
	on the evolution of the orbit is thus taken into account as well.

\subsection{Initial distribution of orbital periods and separations}
\label{sub:initial}
    
    \begin{figure}[!t]
        \includegraphics[width=0.50\textwidth]{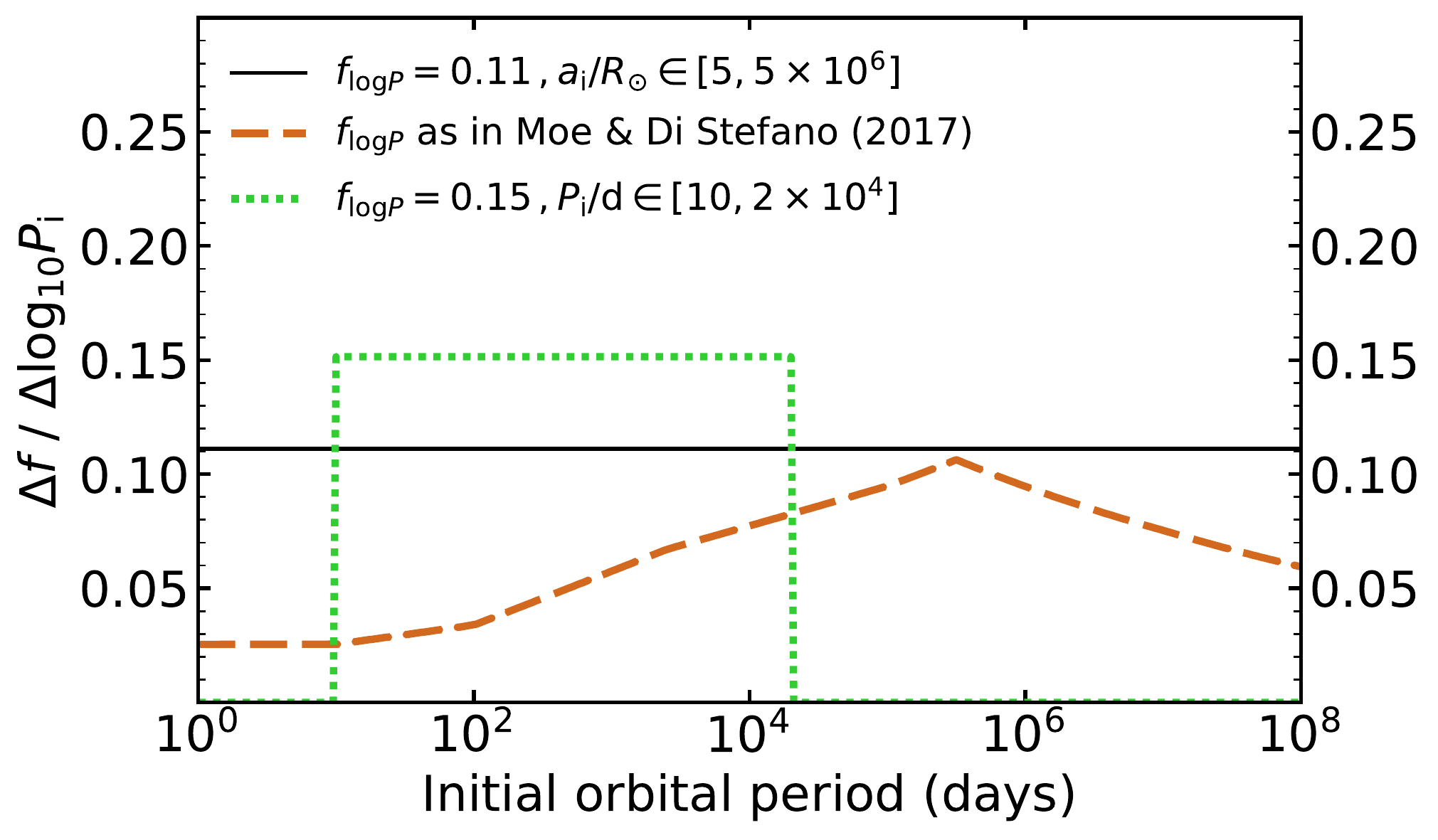}
	    \caption{Initial period distributions adopted in our models. The vertical axis represents the binary fraction in each period bin,
	    		i.e. $\flogP$. The black-solid line represents our default initial distribution of separations, which is flat in $\log_{10} \sepi$
	    		between $5\Rsun$ and $5\times10^6\Rsun$, with $\flogP=0.11$ over this interval. The brown-dashed line shows the
	    		prescription of \cite{Moe2017}, adopted in set M13, for a $1.0\Msun$ primary and mass ratio $\Msec/\Mprim>0.1$.
	    		The shape and maximum of the distribution of \cite{Moe2017} depend on the primary mass and mass ratio.
	    		The green-dotted line shows a period distribution flat in $\log_{10} \Pin$ and with $\flogP=0.15$ in the period range
	    		$[10,2\times10^4]$ days, which we adopt in our model set M14. The integral of each curve represents the total
	    		binary fraction over the period interval.
	    		}
    \label{fig:Pini}
    \end{figure}
    
   In our model sets it is assumed by default that the initial distribution of separations is flat 
    in $\log \sepi$ over the range $[5,\,5\times10^6]\,\Rsun$, with a constant binary fraction per 
    decade of orbital period, $\flogP$, of approximately $0.11$ in this interval.
    This choice has the advantage of being easy to implement and to compare with previous results of population synthesis studies. 
    Furthermore, it is broadly consistent with the observed orbital separations of binary systems in the young 
    stellar association Scorpius OB2 \cite[][]{Kouwenhoven2007} and with the data of \citet[][Fig.~37]{Moe2017} 
    for the orbits of A/late-B-type binaries with primary masses in the range $2$--$5\Msun$. However, solar-type 
    stars with masses between $0.8$ and $1.2\Msun$, which are the most frequent primary masses of the progenitor
    systems of our synthetic CEMP-$s$ stars, have a rather different orbital-period distribution and a lower 
    overall binary frequency \cite[][]{Raghavan2010, Moe2017}. In addition, by default we assume in our models that the
    initial distributions of period and mass ratios are independent, hence the joint probability
    of forming a binary system with initial period $P$ and mass ratio $Q$, 
    $p(P, Q)$, is the product of the individual probabilities $p(P)$ and $p(Q)$.
    By contrast, \cite{Moe2017} find that the period and mass-ratio 
    distributions of observed binary stars are not separable, but closely interconnected. \cite{Moe2017}
    determine a set of equations to calculate the joint probability function of forming 
    a binary system with primary mass $M_1$, mass ratio $Q$, and orbital period $P$. 
    We implement this set of equations in our model set M13.
    
    Anticipating the results of Sect. \ref{results}, our simulations in general form CEMP-$s$
    stars at much longer orbital periods than observed, unless angular momentum is removed
    from the system with extremely high efficiency. For the sake of comparison, in our model set
    M14 we assume that all binary systems are formed with orbital periods in the range $[10,2\times10^4]$ days,
    which is approximately the range in which the CEMP-$s$ stars are observed by \cite{HansenTT2016-2},
    and that the total binary fraction is $50\%$ over this interval. 
    Although we have little information about the initial binary and orbital properties of very metal-poor
    Halo stars, in particular for relatively wide binaries, the resulting value of $\flogP=0.15$ is in 
    approximate agreement with the results of \cite{Gao2014} for metal-poor F/G/K stars in binary systems
    with $P \la 1,\!000$~d. %
    Figure \ref{fig:Pini} shows the three different initial period distributions adopted in our simulations.

\subsection{Detection probability of the orbits.}    
\label{sub:detect}

	In order to compare our simulations to the observed distribution of systems, we need to take account of the likelihood 
	of a given synthetic binary to be detected by the observing campaign of \cite{HansenTT2016-2}, whose strategy consisted 
	of taking observations roughly every $30$ days for around $3,\!000$ days. We compute this likelihood using a 
	Monte Carlo method. For a given set of system parameters (primary and secondary masses, orbital period and eccentricity), 
	we randomly select an angle between the orbit's major axis and the line of nodes, a value for the cosine of the inclination of the orbital plane of
	the binary to the plane of the sky%
	\footnote{For random orientations of the orbit, the cosine of the inclination angle is uniformly distributed.}, %
	and a starting point in the orbit. 
	We compute the line-of-sight velocity at this point and 
	at every $30$ days for $3,\!000$ days, recording the maximum and minimum velocity. The difference between these is compared 
	to the threshold radial-velocity amplitude of the observations, and if it is above this threshold, 
	the system is deemed to have been detected. In their study, \cite{HansenTT2016-2} achieve a $0.1\kms$ precision
	in their radial-velocity measurements, and we consider this value to be our detection threshold, $\Kmin$.

    \begin{figure}[!t]
        \includegraphics[width=0.5\textwidth]{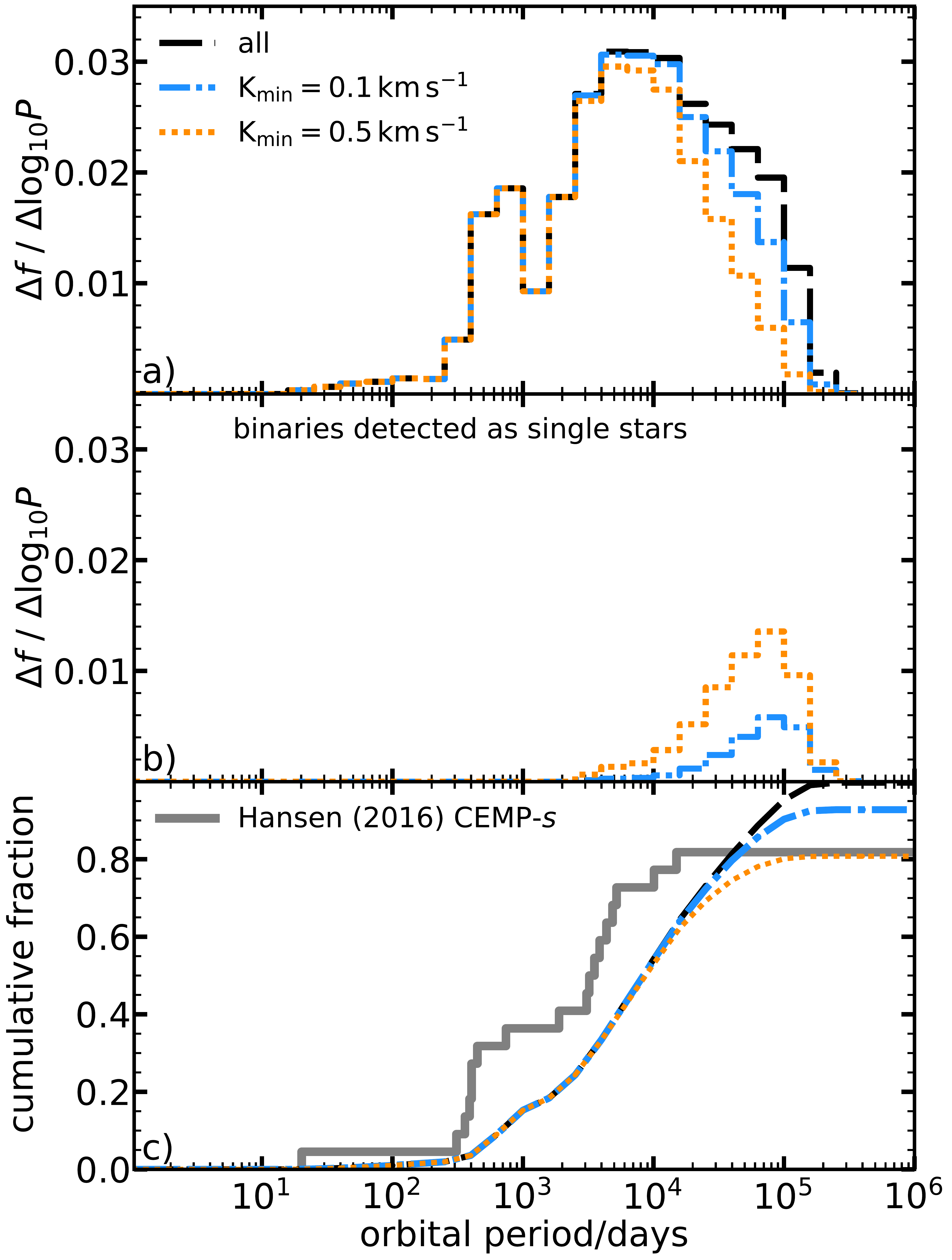}
	    \caption{(a) Period distributions of synthetic CEMP-s stars computed with our model set M2
	    		and with different adopted detection thresholds for radial-velocity variations. 
	    		The black-dashed line shows the period distribution of all the CEMP-$s$ stars in our simulation.
	    		The blue-dot-dashed and orange-dotted lines show the period distributions of the simulated CEMP-$s$
	    		stars with detection thresholds of $K_{\mathrm{min}}=0.1$ and $0.5\,\km\,\s^{-1}$, respectively.
	    		(b)~Same as in panel (a) for synthetic CEMP-$s$ stars that would be
	    		detected as singles with the above thresholds.
	    		(c)~The cumulative orbital-period distributions corresponding to the models
				in panel (a) are compared with the observed distribution calculated with the data of 
				\citet[][grey-solid line]{HansenTT2016-2}.
				}
    \label{fig:detectability}
    \end{figure}

	We repeat this for $10^5$ choices of the orientation of the major axis, the inclination, and starting point in the orbit. 
	The detection probability is then given by the number of systems that exceed the detection threshold, divided by the 
	total number of iterations. We compute detection probabilities for a grid of potential binaries.
	To limit the number of systems we need to compute, we first determine the relevant parameter range of systems, as explained below.
	We then interpolate in this grid to find the detection probability of any CEMP-$s$ system returned by 
	the population-synthesis calculations.

	In the binary systems of our simulated CEMP-$s$ population, the primary is the carbon-rich star which would be observed. 
	Its mass should be higher than $0.5\Msun$, or its luminosity would be so low ($L_{\star}\lesssim 0.08\Lsun$) that the
	$V$-magnitude would typically not satisfy our selection criteria (see Sect. \ref{sub:popsyn}),
	and lower than about $0.95\Msun$, otherwise the star would have become a white dwarf before ten
	billion years \cite[see e.g. Fig.~7 of][]{Abate2015-3}. The secondary star
	is a white dwarf of mass that depends on the initial progenitor mass, and in our simulations varies mostly in the range $[0.5,0.8]\Msun$.
	CEMP-$s$ systems at periods longer than $10^6$ days are not found in our simulations. 
	For binary systems with periods shorter than $3,\!000$ days the full radial-velocity curve would be sampled 
	by the observing campaign of \cite{HansenTT2016-2}. Such a binary would still go undetected if observed close 
	to face-on ($i\lesssim 1^{\circ}$ for a detection limit of $\Kmin = 0.1~\kms$), but the probability of such 
	an unfavourable inclination is less than about $1\%$. We therefore assume a detection probability of $100\%$ 
	for $P < 3,\!000$~d.
	In conclusion, our grid of detection probabilities covers total system masses in the 
	range $[0.7,2.0]\Msun$, secondary masses in the range $[0.5,1.0]\Msun$, and orbital periods in 
	the range $[10^3,10^6]\,\days$. Because all our synthetic systems are circular, we do not account
	for the detectability of eccentric orbits (but see the discussion in sections \ref{sub:obs} and \ref{sub:ecc}).
	
	Figure \ref{fig:detectability} illustrates the effect of different radial-velocity detection thresholds
	on the period distribution of the synthetic CEMP-$s$ systems computed with our default model set M2. 
	In the top panel of Fig.~\ref{fig:detectability}, the black-dashed line shows the differential period
	distribution of all our synthetic CEMP-$s$ stars. The  blue-dot-dashed and orange-dotted lines show the
	CEMP-$s$ stars that would be detected as binary systems with threshold radial-velocity amplitudes of
	$0.1$ and $0.5\,\km\,\s^{-1}$, respectively. %
	Figure~\ref{fig:detectability}b shows the period distributions of CEMP-$s$
	systems that would be detected as single stars with the observation strategy and thresholds described
	above. In Fig.~\ref{fig:detectability}c the same three models as in panel (a) are compared to the observed cumulative period distribution (grey-solid line).
	Only $18$ out of the $22$ CEMP-$s$ stars observed by \cite{HansenTT2016-2} ($\approx 82\%$ of the sample)
	are confirmed binaries, $17$ of which have a determined period while the binary with an as yet undetermined
	period is tentatively plotted at $P = 15,\!000$\,d in Fig.~\ref{fig:detectability}. We make the assumption
	that the other four stars also belong to binary systems but have periods too long to be detected 
	(indicatively $\Porb>15,\!000$ days). As expected, the proportion of CEMP-$s$ stars detected as
	binary systems in our simulations decreases with increasing radial-velocity threshold $\Kmin$.
	A binary fraction among simulated CEMP-$s$ stars consistent with the observations is found 
	adopting $\Kmin=0.5\,\km\,\s^{-1}$ in our model set M2. Nonetheless, to be consistent with
	the precision achieved in the work of \cite{HansenTT2016-2} we adopt $\Kmin=0.1\km\,\s^{-1}$.

\section{Results}
\label{results}

	We evolve a population of very metal-poor binary stars for each set of initial assumptions
	described in the previous sections. We select the CEMP-$s$ stars and we calculate the 
	orbital-period distribution for these systems, which we subsequently compare 
	with the observations of \cite{HansenTT2016-2}. 
	Columns 7--9 of Table~\ref{tab:all_models} characterise the resulting period 
    distribution of each model set by providing the logarithmic orbital periods at 
    $2.5$, $50$ and $97.5$ percentiles of the synthetic CEMP-$s$ population
    (i.e. the orbital period at which the cumulative period distribution is equal 
    to $0.025$, $0.5$ and $0.975$, respectively).
    
    For each model set, we perform a Kolmogorov-Smirnov (K-S) test to
    evaluate the likelihood that the observed period distribution is drawn
	from the corresponding synthetic distribution. Column 10 in Table~\ref{tab:all_models}
	shows the resulting \emph{p}-values%
	\footnote{These were calculated with a version, adapted for \texttt{Python}, of the 
	procedure \texttt{ksone} presented in ``Numerical Recipes'' \cite[][]{NumericalRecipes}.}, %
	which give an indication of the relative goodness-of-fit of each model. We note that eleven of our 
	model sets have $p$-values less than $0.05$, which is the threshold often used as a criterion to 
	reject a model with statistical significance. While this result suggests that most of our models are
	incompatible with the observed period distribution, in the following we will discuss in detail how different 
	model assumptions modify the theoretical orbital-period distributions and why many of these fail to reproduce the data. %

	All figures in this section consist of two panels, as in Fig.~\ref{fig:detectability}. 
	In the top panel we show the differential period distributions predicted with our models,
	before accounting for the detection probability of the orbit. 
	The results are normalised such that the integral of each curve is equal to the total 
	CEMP fraction computed with that model, $\fC$, which is reported in the top-right corner of the plot
	and in Table~\ref{tab:all_models}. The bottom panels show the corresponding cumulative period distributions, 
	after applying a radial-velocity detection threshold of $\Kmin=0.1\km\,\s^{-1}$. These are compared to 
	the observed cumulative period distribution, shown as a thick, solid grey line. Each cumulative 
	distribution is normalized to the total CEMP-$s$ population, either observed or modelled, such that 
	the value at $P = 10^6$\,days corresponds to the detected (or detectable) binary fraction in the population.
	Our default model set M2 is always shown as a reference with a black-solid line.

\subsection{Changes in the stability criterion of Roche-lobe overflow}
\label{res:RLOF}

\begin{figure}[!t]
        \includegraphics[width=0.5\textwidth]{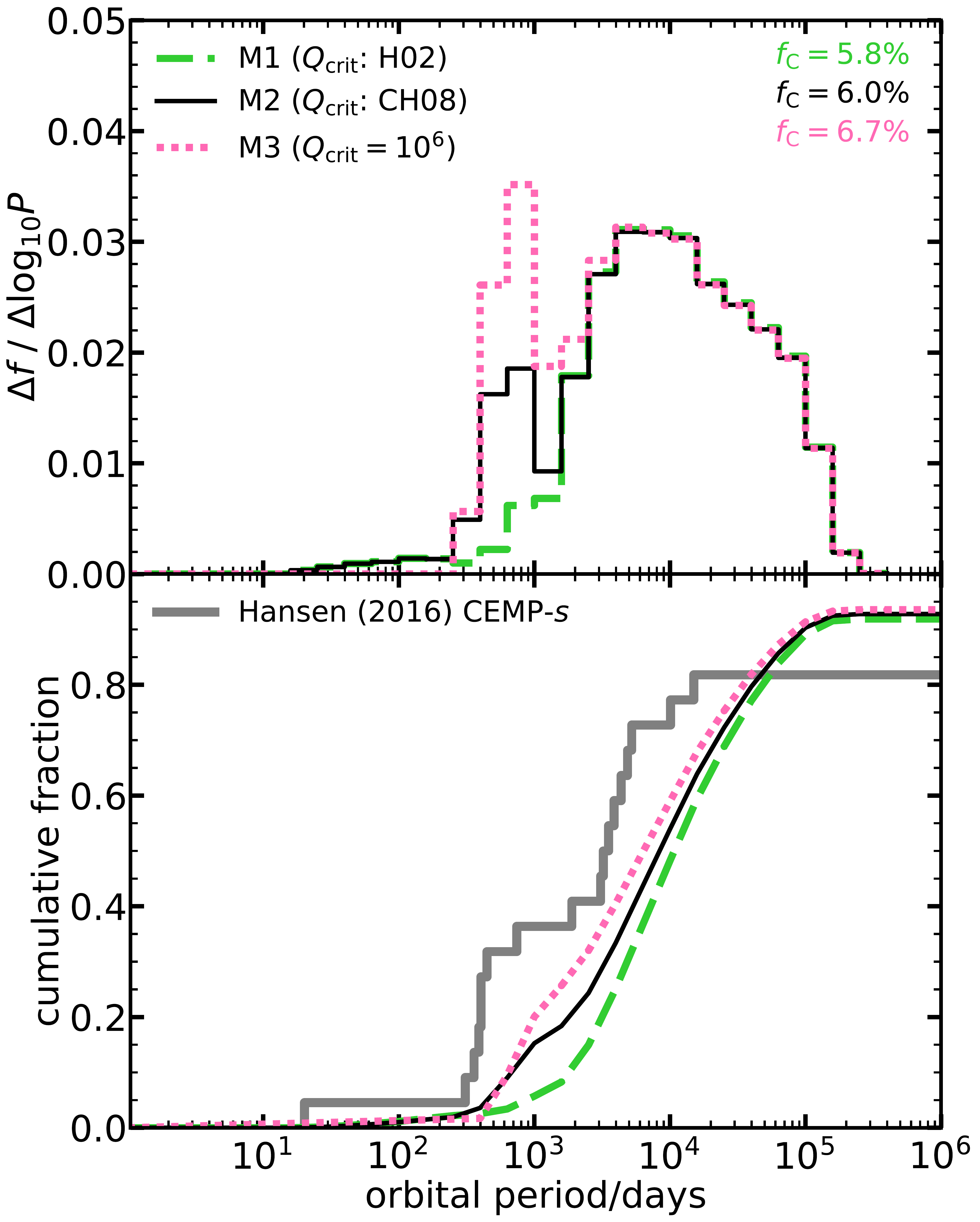}
        \caption{As Fig.~\ref{fig:detectability} for models with different RLOF stability criteria.
        		 Model set M1 adopting the H02 criterion is shown by the green-dashed line. 
        		 Model set M2 (thin, black-solid line) uses the CH08 criterion. 
        		 Model set M3 with $\qcrit=10^6$ (i.e. RLOF from AGB stars is always stable) is indicated by the magenta dotted line.
        		 The top panel shows the entire simulated populations, while in the the bottom panel only
        		 CEMP-$s$ stars detectable as binary systems are shown ($\Kmin=0.1\km\,\s^{-1}$ is adopted).
        		}
    \label{fig:qcrit}
    \end{figure}

	Figure~\ref{fig:qcrit} shows the period distributions of the models sets M1, M2 and M3, 
	with different stability criteria for RLOF from AGB donors.
	As expected, if the value of $\qcrit$ increases, meaning that RLOF is stable for a larger range
	of binary systems, the number of CEMP stars with periods between approximately a few hundred 
	and a few thousand days is increased. This is roughly the interval at which the primary stars
	avoid filling their Roche lobes on the first giant branch so that the RLOF phase can occur when
	the donors have reached the AGB. In this period range we find systems in which the secondary star 
	accretes enough material to become carbon-enriched through stable RLOF.
	This is best seen in the top panel of Fig.~\ref{fig:qcrit}, where 
	the distributions of sets M2 and M3 have a peak at periods between about $300$ and $1,\!000$~days, 
	as a result of the increased stability of RLOF.
	
	The proportion of CEMP stars with periods less than $2,\!500$
	days is
	$15\%$ and $24\%$
	in model sets M1 and M2, respectively, whereas it is $32\%$
	in set M3 in which RLOF is always stable.
	This implies that at least two thirds
	of our synthetic CEMP stars, most realistically more, are formed by accretion of stellar
	winds. Were RLOF the only efficient mechanism to transfer material in low-mass binary stars, 
	the fraction of CEMP stars in our metal-poor population would therefore be at most $2\%$, 
	even if RLOF is always stable, that is, at least a factor of three lower than the fraction 
	determined from the observed SDSS/SEGUE sample ($\approx~\!6.1\%$ for stars with $[\Fe/\Hy]\approx-2.0$, \citealp{Lee2013}).
	
	In addition, Fig.~\ref{fig:qcrit} shows that at any value of the cumulative fraction the synthetic
	distributions overestimate the observed periods approximately by a factor between $2$ and $10$.
	The mismatch is also reflected in the small $p$-values in Table~\ref{tab:all_models}. 
	While set M3 has a marginally acceptable $p$-value of $0.05$, this model lacks physical realism.
	Furthermore, a large proportion of simulated CEMP stars have periods above $15,\!000$ days,
	most of which should be detectable with the observing strategy of \cite{HansenTT2016-2}.
	In their sample, only one CEMP-$s$ star out of 22 (about 5\%) is detected as a binary with 
	an as yet undetermined period that is presumably at least $15,\!000$ days, while four stars
	(about 18\% of the sample) have apparently constant radial velocity.
	By contrast, the proportion of detectable synthetic CEMP binaries with periods above $15,\!000$ days is about $26\%$ 
	in model set M3, and it is $29\%$ and $32\%$ in the more realistic sets M2 and M1, respectively.
	The expected fraction of undetected binaries, with radial-velocity amplitude less than $0.1\,\kms$, is only about $7$\% for all three models.
	This confirms that, regardless of the assumptions about RLOF stability, a significant fraction of systems
	form at very long separations. We therefore conclude that, while a better understanding of the RLOF process is necessary 
	to reproduce the proportion of observed CEMP-$s$ systems with periods up to a few thousand days, it is
	not sufficient to solve the discrepancy between models and observations.
	To reproduce at the same time the fraction of observed CEMP-$s$ stars and their period distribution
	it is necessary to correctly account for wind mass transfer.

\subsection{Varying the accretion efficiency of wind mass transfer}
\label{wind-accretion-efficiency-model}

	\begin{figure}[!t]
	\includegraphics[width=0.5\textwidth]{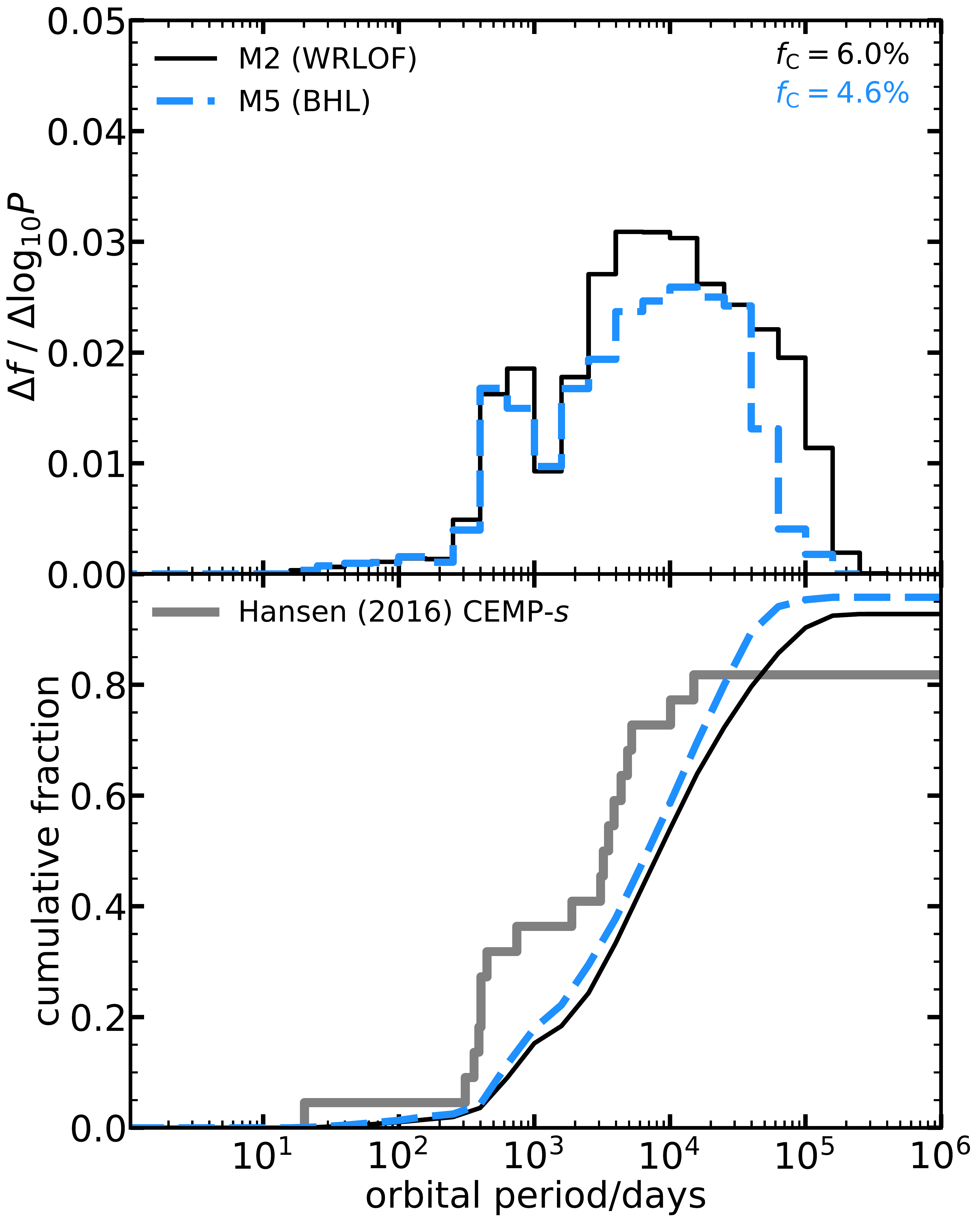}
	\caption{As Fig. \ref{fig:qcrit} for model sets M2 and M5 (dashed-blue line) in which
			 the wind accretion efficiency is computed with the WRLOF and BHL
			 prescriptions, respectively.
			}
	\label{fig:wind1}
	\end{figure}
	
	Figure~\ref{fig:wind1} compares the period distributions of model sets M2 and M5 (dashed-blue line), 
	in which the wind accretion efficiency is computed with the WRLOF and BHL models, respectively. The WRLOF
	model predicts higher accretion efficiencies than the BHL prescription over a large range of separations, also in
	wide systems \cite[][]{Abate2013}. Consequently set M2 produces CEMP stars at longer periods than set M5. 
	However, the BHL model set M5 is only marginally closer to reproducing the observed period distribution than the WRLOF model.
	Furthermore, this comes at the expense of a predicted CEMP fraction of $4.6\%$, which underestimates the 
	results of the SDSS/SEGUE survey \cite[$\approx 6.1\%$][]{Lee2013}, 
	and a predicted fraction of undetected binaries of only about $4\%$, even lower than in model set M2.

\subsection{Changes in the adopted angular-momentum loss}
\label{amloss}

	\begin{figure}[!t]
	    \includegraphics[width=0.5\textwidth]{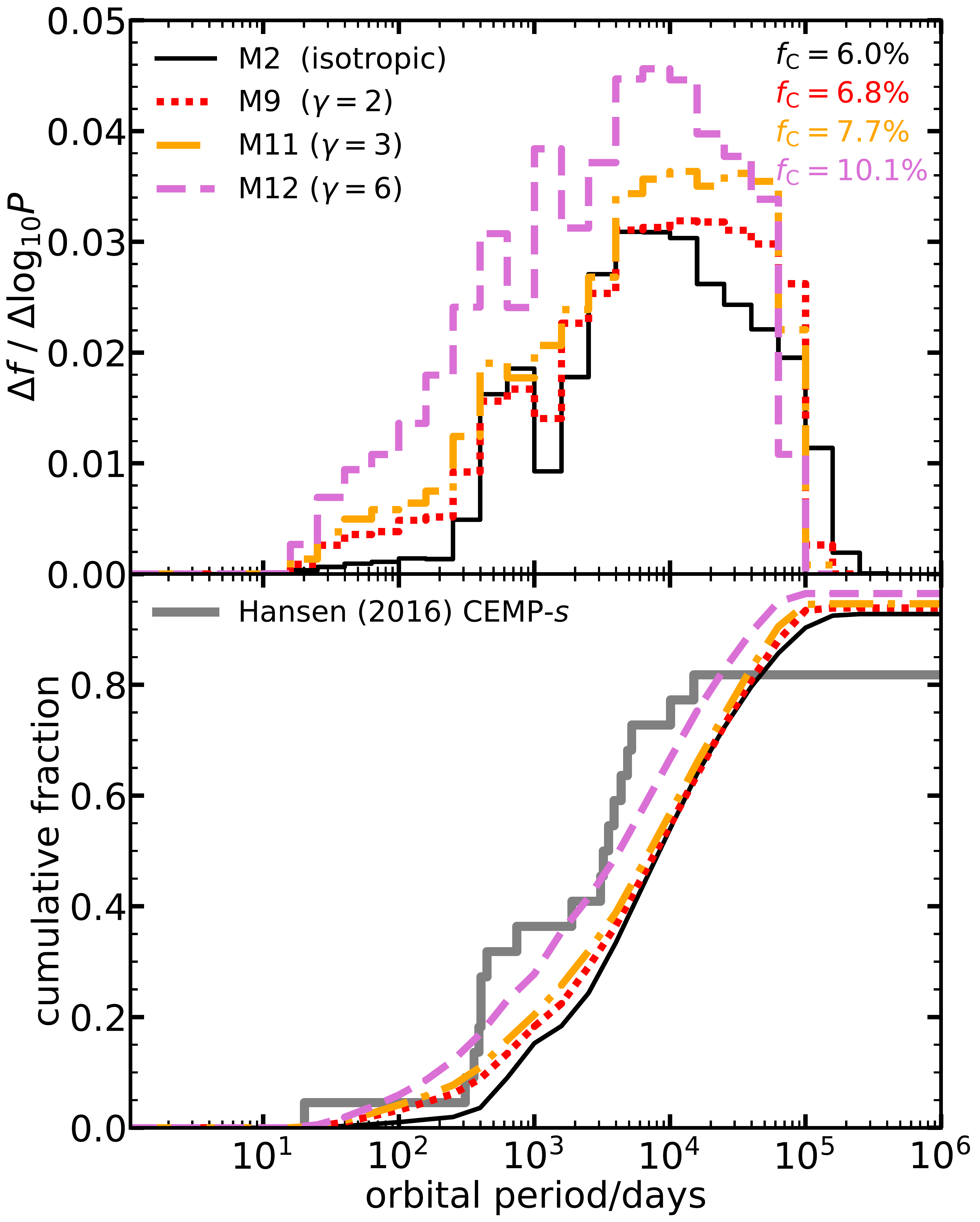}
	    \caption{As Fig. \ref{fig:qcrit} for different for angular-momentum-loss prescriptions.
	    		 Model set M2 assumes isotropic wind mass loss.
	    		 Model sets M9, M11, and M12 (red-dotted, orange-dot-dashed and violet-dashed lines, respectively) are computed
	    		 with Eq. (\ref{eq:gamma}) and $\gamma=2,3,10$, respectively.
	    		}
	\label{fig:amloss1}
	\end{figure}

\begin{figure}[!t]
	    \includegraphics[width=0.5\textwidth]{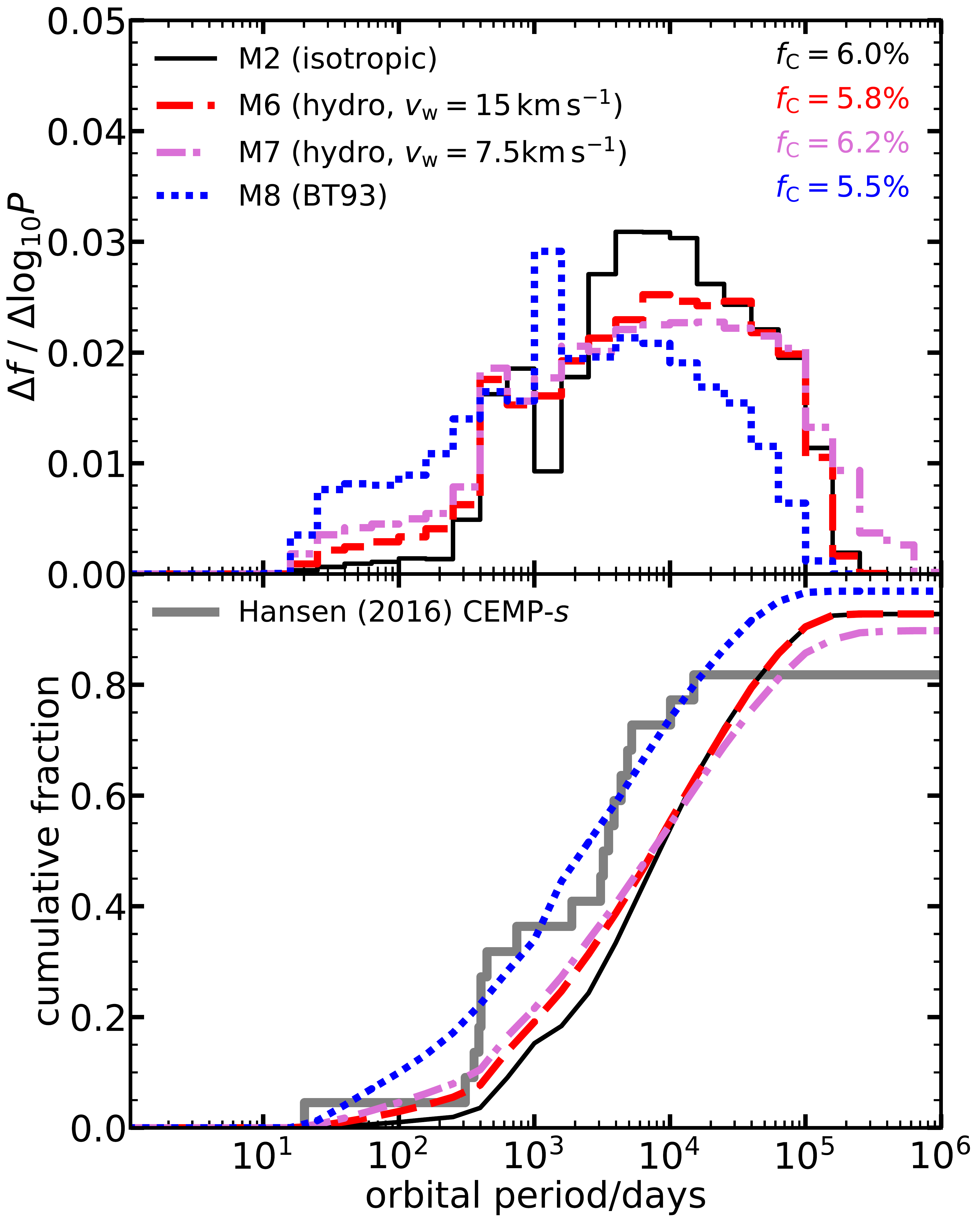}
	    \caption{As Fig. \ref{fig:amloss1} with model sets M6 (dashed line) and M7 (dot-dashed line), in
	    		 which the angular-momentum loss is computed with Eq. (\ref{eq:hydro}) and $\vw=15$ and $7.5\km\,\s^{-1}$,
	    		 respectively, and set M8 (dotted line), in which Eq.~(\ref{eq:BT93}) is adopted.
	    		}
	\label{fig:amloss2}
	\end{figure}

	Figures \ref{fig:amloss1} and \ref{fig:amloss2} show the period distributions obtained with
	model sets adopting different assumptions about the angular-momentum loss. In all models shown,
	the WRLOF prescription of wind accretion efficiency and the CH08 criterion of RLOF stability
	are used.
	Model sets M9, M11 and M12 in Fig. \ref{fig:amloss1} are computed using $\gamma=2,3,6$ in Eq. (\ref{eq:gamma}),
	whereas set M2 assumes the wind is expelled isotropically by the binary system. The CEMP fraction
	increases with increasing $\gamma$ because the enhanced angular-momentum loss causes the binary systems to
	shrink more and therefore the range of separations at which the stars can interact is larger.
	Also the number of systems that produce a CEMP star after experiencing a common envelope increases with $\gamma$.
	Many of these had a relatively large initial separation, and hence the accretor had the time
	to accrete material and become carbon-rich before the onset of unstable RLOF.
	These systems appear at $P \lesssim 400$\,days in Fig.~\ref{fig:amloss1}.
	
	For $\gamma=2$ and $3$, the increased angular-momentum loss compared to the isotropic wind model
	does not correspond to a significant shift of the period distributions towards shorter periods.
	To understand this result it is convenient to subdivide the entire range of orbital periods
	of the synthetic CEMP populations into smaller intervals and subsequently compare the initial
	periods of the progenitor binary systems which, with different model sets, end~up in the same
	interval. This exercise shows, for example, that model sets M2, M9 and M11 form the same 
	proportion (approximately $36$--$38\%$) of CEMP stars with orbital periods between $10^3$ and $10^4$
	days, but the progenitor binary systems in the three model sets had different initial-period ranges.
	In the isotropic-wind assumption (set M2) CEMP stars come from systems that had initial periods
	in the interval $1000$--$8000\,\days$. Using Eq. (\ref{eq:gamma}) with $\gamma=2$ (set M9) the
	progenitor binary systems had initial periods mostly between $2000$ and $20,\!000$ days, with
	a tail up to about $10^5$ days. With model set M11 ($\gamma=3$), the initial periods of these
	CEMP stars span between about $2000$ and $50,\!000$ days, with a tail up to a few hundred
	thousand days. 
	
	With model set M12 the cumulative period distribution of CEMP stars is roughly consistent with
	the observations (\emph{p}=0.28, see also Fig.~\ref{fig:amloss1}) because of the combined effect of a strong angular 
	momentum loss by stellar winds and the increased number of systems undergoing a common envelope.
	It should be remembered, however, that the assumption of a constant $\gamma$ is not supported by
	physical arguments and it is unrealistic. In fact, it implies that the specific angular momentum
	expelled by the binary system does not depend on the masses of the stars
	and their distance, which is at odds with the results of hydrodynamical simulations
	\cite[e.g.][]{Jahanara2005, ChenZ2018, Saladino2018-1}. Considering for example a $1\Msun$ primary
	star and a $0.6\Msun$ companion in a $10^5$-day orbit, with $\gamma=6$ the ejected material has 
	a specific angular momentum ten times higher than in the isotropic-wind approximation, despite the
	fact that at wide separations the outflow from the donor star is expected to be essentially spherical 
	\cite[e.g.][]{Shazrene2011, Shazrene2012}.
	
	The distributions computed with the orbit-dependent angular-momentum loss prescriptions of model sets
	M6, M7 and M8 are shown in Fig. \ref{fig:amloss2} with dashed, dot-dashed, and dotted lines, respectively. 
	Model M6 uses Eq. (\ref{eq:hydro}) based on hydrodynamical simulations
	from the literature and predicts an increased proportion of CEMP-$s$ stars at periods shorter than about 
	$2,\!500$ days compared to default set M2, at the expense of systems with periods between about $2,\!500$ 
	and $30,\!000$ days (see top panel of Fig.~\ref{fig:amloss2}). Binaries with $P \lesssim 30,\!000$\,d lose 
	more angular momentum than in the isotropic-wind model and thus evolve into closer orbits. This results in 
	a larger fraction of CEMP-$s$ stars formed through stable RLOF (with $300\,\mathrm{d} \lesssim P \lesssim 2,\!500$ days) 
	as well as systems experiencing a common envelope after accreting enough material to become CEMP-$s$ stars 
	(ending up with $P \lesssim 300$\,d). With model sets M2 and M6
	the proportion of CEMP-$s$ stars with periods up to $2,\!500$ days is $24\%$ and $31\%$, respectively.
	At periods longer than about $10,\!000$ days the cumulative distributions of M2 and M6 are essentially identical,
	and the mismatch with the observations discussed in Sect.~\ref{res:RLOF} therefore persists ($p<$0.05, see Table~\ref{tab:all_models}).

	A critical parameter in Eq. (\ref{eq:hydro}) is the terminal velocity of AGB winds, $\vw$, because a 
	lower wind velocity implies that, for the same orbital period, a larger amount of specific angular 
	momentum is carried away by the ejected material. This parameter is uncertain, as observed wind 
	velocities range between a few and a few tens of~$\km\,\s^{-1}$ \cite[e.g.][]{VW93, Danilovich2015, Goldman2017}.
	For the sake of comparison, in model set M7 we assume $\vw=7.5\kms$, which is half of our default value and consistent 
	with the lowest velocity detected by \cite{Goldman2017} among high-luminosity, high-mass-loss stars 
	in the Large Magellanic Cloud. Consequently, in set M7 the proportion of CEMP-$s$ stars
	with periods below $2,\!500$ days increases even more, to about $34\%$. A low $\vw$ also affects
	the number of CEMP-$s$ stars in very wide orbits ($P \gtrsim 50,\!000\,\days$). These systems do not
	enter the WRLOF regime and consequently the mass-transfer efficiency is calculated according to 
	the BHL prescription \cite[Eq.~6 of][]{BoffinJorissen1988}, which is proportional to $\vw^{-4}$ 
	(for $\vorb\le\vw$). With a low $\vw$ also very wide systems accrete enough material to 
	form a carbon-rich star. As a consequence, about $15\%$ of the detectable synthetic CEMP-$s$ stars have
	periods longer than $50,\!000\,\days$ ($4\%$ have $P>10^5\days$), whereas it is only about $10\%$ in model M6 
	($\approx\!2\%$ at $P>10^5\days$).
	Thus, while the correspondence with observations improves somewhat at 
	the short-period end of the distribution, it becomes worse at the long-period end.

	The results of model set M8, based on the ballistic simulations of \cite{Brookshaw1993}, 
	are roughly consistent with the observed cumulative distribution up to about $10^4$ days, 
	although the fraction of undetectable binaries is much smaller than the observed fraction 
	of apparently single CEMP-s stars. This is not surprising 
	because Eq. (\ref{eq:BT93}) predicts large angular-momentum loss for systems with $\vw\le\vorb$, 
	much larger than both the isotropic-wind model and the hydrodynamics-based prescription up to 
	periods of about $50,\!000$ days. However, as we
	mentioned in Sect. \ref{sub:AM}, it should be kept in mind that this model
	overestimates the amount of angular momentum carried by the ejected winds
	and consequently the effect on the period distribution of CEMP stars, 
	because it ignores the effects caused by gas pressure and radiative acceleration.
	Nevertheless, model set M8 is a useful test case to estimate how much angular 
	momentum binary systems would need to lose in
	order to reconcile the results of our simulations with the observed CEMP-$s$ population.
	
    We note that in model sets M6 and M8 the total fraction of CEMP stars ($5.8\%$ and $5.5\%$, respectively)
    is somewhat lower than in set M2 ($6.0\%$). This is because some binary systems that are 
    just wide enough to avoid unstable RLOF and form CEMP stars in the isotropic-wind case, 
    which widens their orbits, instead become tighter with the larger angular-momentum loss
    of model sets M6 and M8. Many of these undergo common-envelope evolution before sufficient 
    chemical pollution of the accretor has taken place. Unlike in the models with constant $\gamma$, 
    this is not compensated by very wide systems evolving into close enough orbits to become CEMP stars.

	In conclusion, Eq.~(\ref{eq:hydro}) used in sets M6 and M7, which derives from the
	results of hydrodynamical simulations, is at present the only prescription for orbital 
	angular-momentum loss with a physical basis. The evidence that, despite the uncertainty
	on $\vw$, these model sets do not reproduce the observed period distribution very well suggests 
	that other physical aspects in our models may have to be reconsidered.

\subsection{Changes in the range of initial periods}
\label{initial_distribution}
\label{vary-ai}    

    \begin{figure}[!t]
        \includegraphics[width=0.5\textwidth]{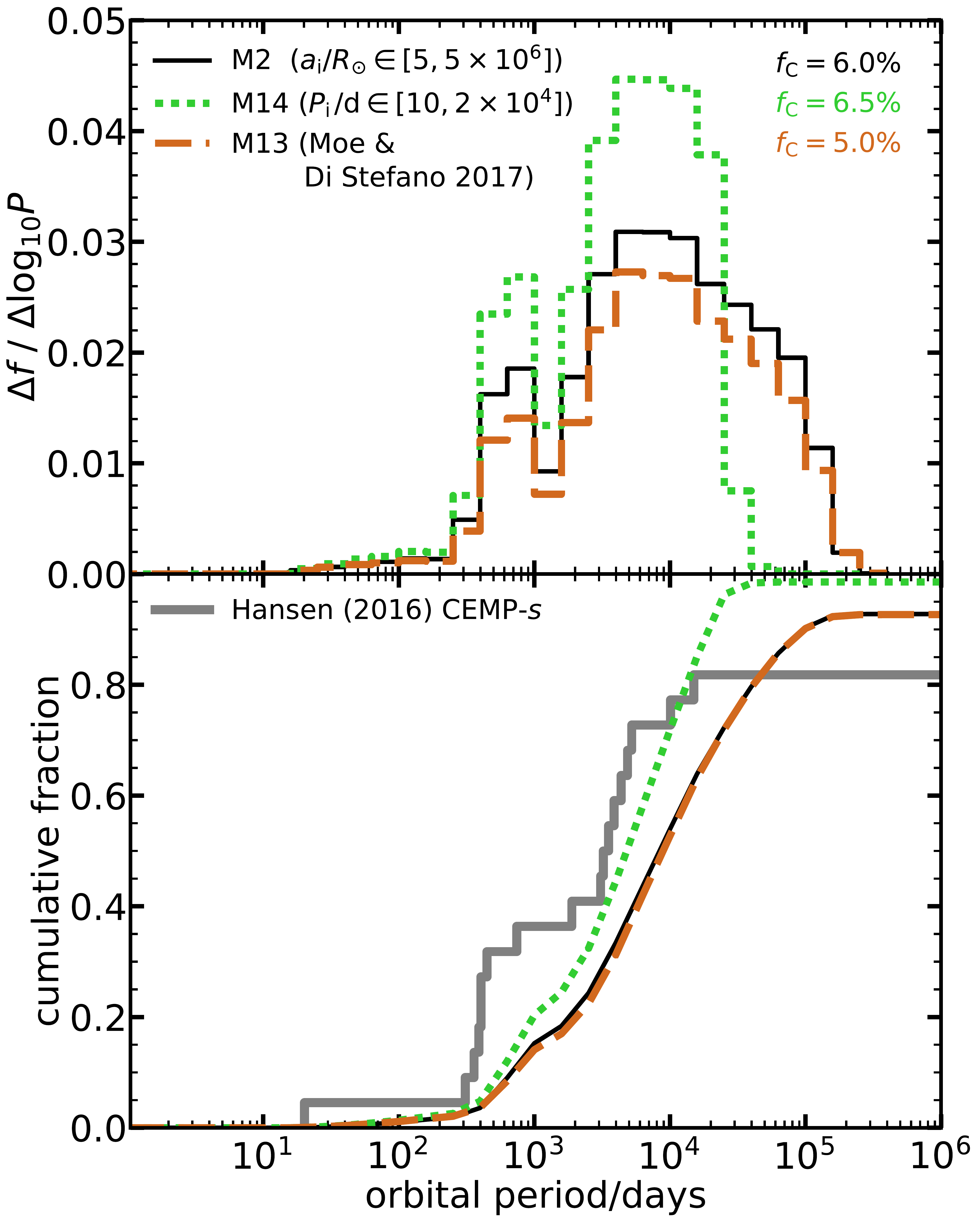}
	    \caption{As Fig. \ref{fig:qcrit} for different distributions of initial separations and periods.
	    		 By default $\sepi/\Rsun$ is in $[5,5\times10^6]$ (black solid line). The dashed line
	    		 is computed using the initial $\log P$--distribution of \citet[][Eq.~23]{Moe2017}. In model set M14
	    		 $\Pin$ is log-flat between $10$ and $20,\!000$ days (dotted line).
	    		}
    \label{fig:ai}
    \end{figure}

    Our model sets M13 and M14 adopt the same assumptions as the default set M2 except for the 
    initial distribution of orbital periods and mass ratios. In model M13 we implement the set of fitting equations 
    to observed binary stars proposed by \cite{Moe2017}. These equations result in a quasi-flat
    initial-period distribution for low-mass primary stars, with a very wide peak between
    approximately $10^4$ and $10^6$ days, and in a combined distribution of periods and mass ratios,
    $q=Q^{-1}=M_2/M_1$, which favours small mass ratios, especially in wide orbits. The  efficiency of
    WRLOF decreases with $q$ in the prescription of \cite{Abate2013}, while it is higher for relatively long-period systems
    around $\approx\!10^4$ days. As a result, these two effects compensate one another and
    the period distribution of CEMP-$s$ stars with model set M13 is very similar to that of
    our default set M2, although with a slightly smaller proportion of systems with $P<10^4\,\days$
    and a decreased overall CEMP fraction.
    
    Our model set M14 has a flat $\log P$--distribution with $\flogP=~0.15$ between $10$ and $20,\!000$ days. 
    These assumptions result in a period distribution which resembles the observations more closely, 
    in particular at the long-period end. By construction, only a small fraction of synthetic 
    CEMP-$s$ stars have periods in excess of $20,\!000$ days, as in the sample of \cite{HansenTT2016-2}. 
    However, this model also predicts that essentially all CEMP-$s$ stars should have detectable 
    radial-velocity variations when monitored with the same strategy and sensitivity as the study 
    of \cite{HansenTT2016-2}, which appears at odds with their finding that four out of $22$ 
    observed CEMP-$s$ stars are apparently single. The \emph{p}-value of $0.11$ nevertheless suggests 
    that we cannot reject this model on statistical grounds. We note, however, that the K-S test is 
    relatively insensitive to differences occurring far from the median of the distribution, as is the case here.

\section{Discussion}
\label{discussion}

\subsection{On the observed binary fraction and orbital periods.}
\label{sub:obs}
	
	We have assumed that all CEMP-$s$ stars are formed in binary systems, a hypothesis borne out by several theoretical 
	and observational studies \cite[][]{Lucatello2005-2, Aoki2007, Bisterzo2012, Lugaro2012, Starkenburg2014}. However,
	four stars in the sample of \cite{HansenTT2016-2} do not exhibit radial-velocity variations consistent
	with orbital motion. Based on the sensitivity of their study, the authors exclude as highly unlikely that all four 
	CEMP-$s$ stars are in binary systems observed face-on, and therefore they conclude that these are single stars.
	However, they did not consider the possibility that these may be binaries with periods much longer than $10^4$ days.
	
	In our simulations we use a Monte Carlo method to account for the likelihood of our synthetic binary systems 
	to be detected with the observing strategy of \cite{HansenTT2016-2}. In their study, the precision achieved 
	in the velocity measurements is about $0.1\kms$. Accordingly, we have assumed a sensitivity to radial-velocity 
	variations of $\Kmin=0.1\km\,\s^{-1}$, and we find a fraction of undetectable binaries which ranges between 
	$2$ and $10\%$ (in model sets M14 and M7, respectively; it is $\approx\!7\%$ in our default set M2).
	This means that, in a sample of 22 CEMP-$s$ stars, we expect between 0.4 and 2.2 undetected binaries, 
	rather than the four that are observed.

	We note that the time span of observations for stars in the Hansen sample is often less than the 3000 
	days we have assumed for computing the detection probability. For two of the four constant-radial-velocity stars 
	it is indeed much shorter: about $1000$ days for HE~$0206$--$1916$ and $800$ days for HE~$1045$+$0226$. It 
	cannot be excluded that evidence for orbital motion would have been detected in these stars if the 
	radial-velocity measurements had lasted for the nominal $3000$ days. %
	In addition, for star HE~$1045$+$0226$ 
	both the errors and the spread in the observed radial-velocity values are substantially larger than 
	$0.1\,\kms$, and an overall downward trend in radial velocities with an amplitude several times $0.1\,\kms$ is 
	compatible with the data (Fig.~1 of Hansen et al.). Several other stars in the sample also have radial-velocity
	measurement errors in excess of $0.1\,\kms$. This suggests that both our assumed time span of 
	$3000$ days and detection threshold of $0.1\,\kms$ are too conservative for the sample as a whole.
	In fact, if we assume a time-span of radial-velocity monitoring of $1,\!000$ days in our simulations,
	the predicted fractions of undetected binaries approximately double. In our default set M2 we expect about
	$16\%$ of undetected CEMP-$s$ systems, in rough agreement with the observed sample (see Fig.~\ref{fig:baseline} in Appendix~\ref{appendixB}).
	In addition, as we show in Sect.~\ref{sub:detect}, if we adopt $\Kmin=0.5~\kms$ in our default set M2
	about $20\%$ of the CEMP-$s$ stars would not be detected as binaries, 
	approximately as in the observed sample. The same is true for most of our simulations,
	with the exception of sets M8, M12 and M14, which would require an even higher detection
	threshold in order to produce approximately $18\%$ of undetected binaries ($\Kmin=1.0~\kms$ in 
	set M12, $\Kmin=1.5~\kms$ in sets M8 and M14).
	
	Perhaps harder to reconcile with our models than the non-detection of radial-velocity variations 
	in four stars, is the paucity of \emph{confirmed} binaries with orbital periods between $10^4$ 
	and $10^5$ days. The Hansen sample probably contains one such a very wide binary system, 
	HE~$0959$--$1424$, for which it was not possible to determine the orbital solution but which 
	exhibits velocity variations of about $2~\kms$. Its orbital period must be longer than 
	$10,\!000$ days, but is likely less than $10^5$ days, and some of the apparently 
	constant-velocity stars discussed above may also turn out to have periods in this range.
	We also note, however, that the smallest velocity amplitude measured in a confirmed CEMP-$s$ binary star 
	is $K=1.57~\kms$, which is significantly higher than our adopted threshold, even assuming a generous value of $\Kmin=0.5~\kms$.
	This is hard to reconcile with a continuous distribution of radial-velocity amplitudes down to the 
	detection threshold, as would be expected from our simulation, and suggests that the paucity of 
	CEMP-$s$ binaries with $P > 10,\!000$~days is real. An approximate upper limit to the period 
	distribution of the order of $10^4$~days is also suggested by the study of \cite{Starkenburg2014}. 
	All of our model sets produce a substantial fraction of detectable binaries with
	periods larger than $10^4$~days. In a sample of 22 CEMP-$s$ stars, between $5$ and $9$
	such binaries are predicted (with $\Kmin=0.1~\kms$). Our only model set that comes close to reproducing the lack of
	systems in this period range is M8 if we make the additional, extreme assumption that $\Kmin=1.5~\kms$, in which case we predict about $2$
	detectable CEMP-$s$ binaries with period $P>10^4$~days. In all other model sets we cannot reproduce at the same
	time the proportion of undetectable binaries and the number of long-period detectable binaries, even if we assume
	a $\Kmin \gg 0.1~\kms$.

	In the discussion above we have only considered circular orbits. With the Monte Carlo method described
	in Sect.~\ref{sub:detect}, we find that eccentric binaries with orbital periods in the range
	$5\times 10^3\lesssim P \lesssim 3\times 10^5~\days$ are less likely to be detected, because the 
	radial-velocity variations are smaller than in the circular case during most of the orbit, except
	when the system is close to periastron. Assuming $\Kmin=0.1~\kms$, the decrease in detection
	probability is about $5\%$ or less for small eccentricities, $e\le 0.3$, while it can be as
	high as about $25\%$ for $e=0.7$, which corresponds to the highest eccentricity determined
	in the observed CEMP-$s$ sample. The smaller detection probability of eccentric systems helps 
	to reduce the difference between the observed and predicted numbers of detectable CEMP-$s$ 
	binaries with $P>10^4$~days, but this effect is too small to remove the discrepancy.
	
	One further aspect that needs to be addressed is whether the observed sample of \cite{HansenTT2016-2}, which we use to constrain
	our models, is representative of the overall CEMP-$s$ population. The members of this sample were initially 
	selected for their chemical properties, namely their observed abundances of carbon
	and barium relative to iron, and subsequently monitored to ascertain whether they belong to binary systems
	and, eventually, to determine their orbital periods \cite[][]{HansenTT2016-2}.
	Consequently, in principle there is no obvious observational bias in favour of relatively short orbital periods.

	Nevertheless, the fact that most stars in the sample have relatively large carbon enhancements ($[\C/\Fe]>1.5$
	in $18$ out of $22$ stars) may introduce a potential bias.
	We note that all sample stars are giants that have already undergone first dredge-up, which has mixed and diluted 
	the transferred material throughout the accretor (with the possible exception of HE$1046-1352$ which has $\logg=3.5$
	and is also the star with the highest carbon abundance, $[\C/\Fe]>3.3$).
	This suggests that they must have accreted substantial amounts of material from their AGB companions
	and, therefore, that the adopted observed sample may be biased towards systems that experienced
	highly efficient mass transfer.	Consequently wide-period systems, which transferred only a few percent of the mass
	ejected by the donor, may be underrepresented. %
	A similar selection effect has been pointed out for barium stars, 
	in which higher $s$-process enrichments are associated with shorter orbital periods \cite[e.g.][]{Boffin1994-1, Boffin2015}. %
	However, if among the CEMP-$s$ stars simulated in model set M2 we select those that have
	$[\C/\Fe]>1.5$ and $\logg \le 3.0$, only very wide systems at $\Porb>10^5$~days are significantly affected.
	These transfer less than a few $0.01\,\Msun$, that is just enough to enhance the surface carbon abundance 
	of the accretor up to $[\C/\Fe]\approx 1$. Because these wide systems make up just a few per cent of 
	the total CEMP population in set M2, the final period distributions
	change only marginally if we exclude them. We conclude that even if there is a bias towards
	large accreted masses, this does not significantly affect the resulting period distribution.

\subsection{Accretion efficiency and angular-momentum loss during wind mass transfer.}
\label{sub:etabeta}

	In the wind mass-transfer process, the amount of mass and angular momentum that are accreted 
	by the companion or lost by the binary system are intricately related. The stronger the wind of 
	the donor interacts with the binary, the higher the accretion rate and angular-momentum loss are 
	expected to be. Many authors have computed hydrodynamical simulations of wind mass transfer in 
	low-mass binary systems with different codes and algorithms, but only a few of these studies have 
	addressed the loss of angular momentum. These simulations produce more or less consistent results 
	concerning the angular momentum lost when similar input physics is adopted \cite[cf. Fig.~11 of][]{Saladino2018-1}. 
	However, the accretion efficiencies found in hydrodynamical simulations differ significantly, 
	by as much as a factor of ten, depending on physical assumptions such as the acceleration mechanism 
	of the wind, the equation of state used to describe the gas particles, the adopted cooling mechanism, 
	and possibly also the algorithms used to compute the accretion rates
	\cite[][]{Theuns1996, Nagae2004, deValBorro2009, Shazrene2010, ChenZ2017, Liu2017, Saladino2018-1}.
	Because of these discrepancies, and with the aim of investigating multiple combinations of model
	assumptions, in our work we chose to treat mass accretion and angular-momentum
	loss as if they were independent processes. 
	To improve on this study it will be necessary to compute these processes self-consistently
	with a model based on a reliable set of hydrodynamical simulations covering a large parameter space.

	We have investigated different models of angular-momentum loss. 
	As demonstrated in Sect.~\ref{amloss}, most of these models yield periods of CEMP-$s$ binaries that
	are significantly longer than observed, with the median of the synthetic period distribution typically
	exceeding the observed value by about a factor of about three.
	With our model sets M6 and M7, which are based on detailed hydrodynamical calculations, the proportion 
	of synthetic CEMP-$s$ stars between $100$ and $2,\!500$ days increases compared to the isotropic-wind model, 
	reducing the discrepancy with the observations in this range. However, at longer periods the ejected 
	material only interacts weakly with the binary and the results are the same as with the 
	isotropic-wind model, hence too many wide-orbit CEMP-$s$ systems are formed.  
	
\begin{figure}[!t]
	    \includegraphics[width=0.5\textwidth]{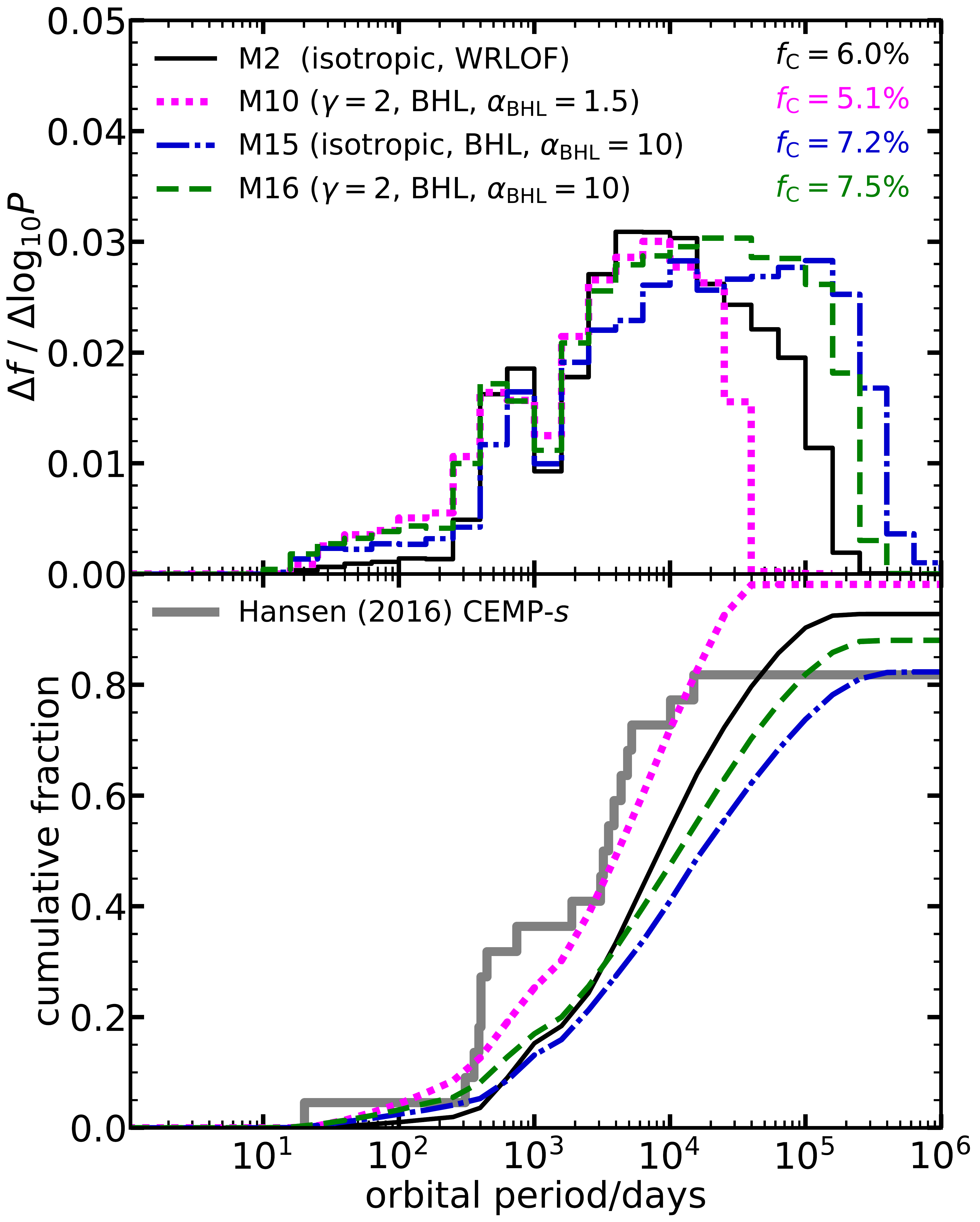}
	    \caption{As Fig. \ref{fig:qcrit} for different for accretion-efficiency and angular-momentum-loss prescriptions.
	    		 Model set M2 adopts the WRLOF model of wind accretion and assumes isotropic wind mass loss.
	    		 Model set M10 (magenta-dotted line) is computed with Eq. (\ref{eq:gamma}) and $\gamma=2$, and adopts the BHL model with $\alpha=1.5$.
	    		 Model sets M15 and M16 (blue-dot-dashed and green-dashed lines, respectively) compute wind accretion efficiency with $\alpha=10$ within the BHL model.
	    		 Set M15 assumes isotropic wind mass loss, whereas set M16 uses Eq. (\ref{eq:gamma}) with $\gamma=2$.
	    		}
	\label{fig:amloss3}
	\end{figure}

	In order to obtain a period distribution of CEMP-$s$ stars that is consistent with the 
	observed population, at least for periods up to about $10^4$ days, it is necessary to assume
	that much larger amounts of angular momentum are lost with the ejected wind material. 
	This is the case for model sets M8, based on the simulations of \cite{Brookshaw1993}, 
	and M12, with $\gamma=6$ in Eq.~(\ref{eq:gamma}), in which binary systems shrink 
	significantly in response to mass loss. However, we emphasize that sets M8 and M12 are 
	not based on realistic physical assumptions. Set M8 is based on ballistic simulations 
	that do not take into account gas pressure and radiative acceleration, which would make
	the ejected outflow more isotropic, and hence the angular momentum dispersed by wind
	material is overestimated. Assuming a constant $\gamma$ in set M12 implies that the
	specific angular momentum of the ejected material is independent of the orbital 
	period of the systems. In addition, we note that with model sets M8 and M12 the
	$20\%$ widest CEMP-$s$ systems have periods mostly in the range 
	$15,\!000$--$40,\!000$ days, of which the majority should be detectable as binaries 
	(however, see the discussion in Sect.~\ref{sub:obs}).
	In their study of the periods and eccentricities of barium stars, \cite{Izzard2010} show that
	they reproduce the bulk of the observed distribution by adopting Eq.~(\ref{eq:gamma}) with $\gamma=2$.
	With the same assumptions, in our model set M10 we find a cumulative distribution approximately consistent
	with the observations for periods up to about $10^4$~days. However, at longer periods set M10 predicts
	only $2\%$ undetected CEMP-$s$ binary systems, and an overall CEMP fraction of about $5\%$, which is
	lower than the observed one (see Fig.~\ref{fig:amloss3}). %

	\cite{Abate2015-1} found that high accretion efficiencies are required to reproduce the surface abundances
	of their sample of observed CEMP binary stars. A similar conclusion was reached by \cite{Abate2015-2},
	who found that approximately $40\%$ of the analysed CEMP-$s$ stars accreted more than $0.1\,\Msun$
	by wind mass transfer from an AGB companion. In their best-fitting model set~B, \cite{Abate2015-1} adopted the 
	equation for the BHL wind accretion efficiency proposed by \citet[][Eq.~6]{BoffinJorissen1988} but they
	arbitrarily replaced the numerical constant $\aBHL=1.5$ with $\aBHL=10$. In their
	model, the high accretion efficiency was combined with strong angular-momentum loss ($\gamma=2$ as in our set M9)
	which was found to be necessary to reproduce the observed orbital periods. %
	As a test, we computed two simulations adopting the BHL prescription for wind accretion
	and $\aBHL=10$, while for the angular-momentum loss either an isotropic wind (set M15) or 
	Eq.~(\ref{eq:gamma}) with $\gamma=2$ (set~M16) are assumed. The results are shown in Fig.~\ref{fig:amloss3}.
	The increase in accretion efficiency generated by $\aBHL=10$ causes a shift in the period 
	distribution of CEMP-$s$ stars towards longer periods than in model sets M4 and M5, %
	which use the BHL model with $\aBHL=1.5$. This is because in these two sets most systems with 
	periods longer than about $60,\!000$ days do not transfer enough mass to generate CEMP stars, 
	whereas they do if we adopt $\aBHL=10$. As a result, about $35\%$ of the whole synthetic CEMP-$s$ 
	population come from very wide systems with periods between $60,\!000$ days and a few times 
	$10^5$ days. In model set M16, which has
	the same assumptions of set B of \cite{Abate2015-1}, the cumulative distribution we determine
	is shifted towards periods a factor of two up to ten longer than in our set M2.
	Overall these results indicate that a simple increase in wind accretion efficiency and specific angular-momentum
	loss of the ejected material applied to all systems regardless of their separations aggravate
	the discrepancy between synthetic and observed period distributions. %
	
	In our simulations, we implicitly ignore that the material transferred to the secondary star carries angular
	momentum, which will spin up the accretor. If the angular momentum content is too great, it can
	prevent the accretion of material \cite[][]{Packet1981}. \cite{Matrozis2017-1} showed that the
	transferred material has to dissipate most of its angular momentum for the secondary
	star to accrete more than a few $0.01\Msun$. Investigating this issue and the constraints
	it puts on the mass-accretion process is beyond the scope of this paper.

\subsection{On the initial orbital-period distribution.}
\label{sub:Pi}

	The distribution of initial orbital periods (or separations) and the initial binary fraction are 
	among the largest uncertainties in our simulations. \cite{Moe2017} have combined and integrated 
	the formidable efforts made by many authors to characterise the orbital-period and mass-ratio 
	distributions of young main-sequence binary systems in the Galactic disk. Adopting their fitting
	equations to these distributions instead of our default $\log a$-flat initial distribution in our
	simulations does not significantly change the period distribution of our synthetic CEMP-$s$ stars
	(see our set M13, Fig. \ref{fig:ai}). 
	
	The study of \cite{Moe2017} does not include a dependence on the metallicity. 
	In particular, little is known about the initial binary properties
	of the low-metallicity Halo population \cite[see e.g.][and references therein]{Duchene2013}.
	For example, \cite{Rastegaev2010} and \cite{Hettinger2015} argue that the total binary fraction increases with
	metallicity. In contrast, \cite{Gao2014, Gao2017} and \cite{Badenes2018} 
	find an anticorrelation between these two quantities. In addition, \cite{Rastegaev2010} has determined
	the orbital-period distribution of a sample of about $60$ Population-II subdwarf binaries.
	The result is an asymmetric distribution with a broad peak between periods of $10$ and $10^4$~days
	and a tail up to about $10^{10}$~days \cite[see Figs.~8 and 10 of][]{Rastegaev2010}.
	This distribution is not dissimilar to what we assume in our model M14, although the $\flogP$
	within the peak of their distribution is smaller, about $0.10$. 
	The result of set M14 is a cumulative period distribution of CEMP-$s$ stars that
	resembles the observed distribution 
	for periods up to about $15,\!000$ days. In addition, with $\flogP=0.15$ we find a CEMP fraction
	of $6.5\%$, similar to that determined from SDSS data \cite[$\approx\!6\%$,][]{Lee2013}.
	We emphasize that the initial-period distribution in set M14 is adopted for comparison purposes and
	it is not claimed to be realistic. However, this choice is broadly in agreement with the study of \cite{Gao2014} for F/G/K stars at 
	$[\Fe/\Hy]<-1.1$ in binary systems up to one thousand days, and it is also roughly consistent
	with the results obtained by \cite{Moe2017} for relatively massive stars ($M_1\gtrsim 5\Msun$), 
	although these are a marginal fraction of all CEMP-$s$ progenitors in our simulations.
	Another source of uncertainty is the 
	contribution to the observed period distribution from binary systems with a white dwarf as a companion.
	The previous evolution of these systems has probably modified their orbit, and consequently their current periods
	differ from the initial periods at their formation. While in their equations 
	\cite{Moe2017} take into account this effect, it is not discussed by \cite{Rastegaev2010}.

	These results are instructive to understand by how much it is necessary to modify the initial 
	period distribution to reproduce the observed CEMP-$s$ period distribution, but we stress that
	the assumptions adopted in model set M14 are completely arbitrary. 
	In principle, the validity of these assumptions can be tested on a number of different astrophysical phenomena.
	For example, if the binary fraction per decade of periods, $\flogP$, increases at low metallicity 
	for orbits up to about $10^4$ days, the rate of Type Ia supernovae is expected to increase, because these supernovae
	progenitors are formed at periods shorter than about $5,\!000$~days \cite[e.g.][]{Claeys2014}.
	A similar argument applies to close symbiotic binaries and blue stragglers, which are expected to
	be more numerous among metal-poor stars if $\flogP$ is weighted towards short orbits.
	
	In conclusion, it is clear that the initial binary population in the Galactic halo needs 
	to be much better characterised in order to put firmer constraints on the physical processes
	playing a role in the formation of CEMP-$s$ binaries. 
	The final data release of the Gaia mission, in combination with radial-velocity monitoring
	surveys, will hopefully provide insight into the properties of the binary population in the
	halo. In particular, binary systems in which both components are low-mass main-sequence stars
	have likely not evolved much over the ten billion years since their formation, and thus the
	initial distribution of orbital periods may be inferred from the study of their current orbital
	properties.

\subsection{On the eccentricities of CEMP-$s$ stars.}
\label{sub:ecc}
	
	In our simulations the binary systems have circular orbits, although about half of the observed CEMP-$s$ stars have 
	non-zero eccentricities up to $e=0.67$ \cite[][]{HansenTT2016-2}. We performed a number of simulations with the model
	sets described in Sect.~\ref{model} and eight values of the initial eccentricity, $e_\mathrm{i}$, 
	uniformly distributed in the range $[0,0.7]$. We find that the period distributions computed in 
	Sect.~\ref{results} do not vary significantly when non-zero eccentricities are considered.
	CEMP stars with periods longer than about $2,\!500$ days are formed at every eccentricity
	in our range, hence reproducing the spread of the observations. The effect of
	tidal friction is increasingly important at shorter periods, because when the primary star ascends 
	the giant branch it fills a significant portion of its Roche lobe. Below $P\approx 1000$~days
	the binary systems experience RLOF, during which the orbit is fully circularised. As a consequence,
	if tides are as efficient as assumed \cite[][]{Hurley2002} and in the absence of a physical mechanism that re-enhances
	the eccentricity, all these short-period synthetic CEMP-$s$ systems are in circular orbits, in 
	contrast with at least four observed CEMP-$s$ stars \cite[two of which are in the sample of][see Fig.~1]{HansenTT2016-2}.
	
	A mechanism that counteracts the tidal forces in binary systems with a giant component close to filling its
	Roche lobe has been invoked to model the eccentricities in a variety of contexts, including post-AGB binary 
	systems	and barium stars.
	The nature of such a mechanism that can enhance the eccentricity is currently unclear. \cite{VanWinckel1995}
	and \cite{Soker2000} suggested that enhanced mass transfer at periastron may counteract the circularisation
	of the orbit. \cite{Axel2008} successfully applied a tidally-enhanced model of mass loss from AGB stars
	to reproduce the eccentricities observed in moderately wide binary systems with a white 
	dwarf and a less evolved companion, such as Sirius. Alternatively, it has been suggested that
	a fraction of the material expelled by the donor star may escape through the outer Lagrangian
	points and form a circumbinary disk. The interaction between this disk and the binary
	may lead to an increase of the system's eccentricity which depends on properties such as 
	the orbital separation, the mass of the disk, its lifetime, viscosity and density distribution
	\cite[e.g.][]{Artymowicz1994, Waelkens1996, Dermine2013, Vos2015}.
	Regardless of its nature, a mechanism that enhances the eccentricity 
	of close systems will necessarily affect the final orbital periods. Consequently, any model promising to
	reconcile the synthetic period distribution of CEMP-$s$ stars with the observations will have to take
	into account a mechanism to counteract circularisation in short-period binaries. 
	
	In our study we focused on binary systems and we did not consider triples. The presence of a third star
	orbiting a binary system may in some circumstances increase its eccentricity, as described by the Kozai-Lidov mechanism \cite[][]{Kozai1962, Lidov1962},
	opposing the effect of circularisation \cite[see e.g.][]{Perets2012}.
	The observed proportion of triple-to-binary systems varies
	between approximately $10\%$ and $20\%$ depending on the sample and on the detection techniques, 
	\cite[e.g.][]{Eggleton2008-2, Rastegaev2010, Tokovinin2014-2, Borkovits2016}. Consequently, we could expect that
	in CEMP-$s$ sample of \cite{HansenTT2016-2} between two and four objects are, or were, in fact triple systems.
	This might help to explain the presence of some of the eccentric CEMP-$s$ stars with periods shorter than $10^3$ days.
	For example, in the eccentric system CS22964--161AB \cite[$P=252$~days and $e=0.66$, from][]{Thompson2008}, 
	which is not included in the survey of \cite{HansenTT2016-2}, both components are CEMP-$s$ stars in the 
	early phases of their evolution, and hence have likely been polluted in the past by a third object
	\cite[see e.g.][]{Thompson2008, Abate2015-2}.

\subsection{On the origin of single CEMP-$s$ stars}
\label{sub:single}

	If the four CEMP-$s$ stars that do not exhibit radial-velocity variations are actually single stars, 
	their formation history has to be understood. Their observed effective temperatures and surface
	gravities are relatively high \cite[see Tab.~6 of][and references therein]{HansenTT2016-2}, which excludes 
	the possibility that they are self-polluted AGB stars. Also, it is very unlikely that they were formed in binary
	systems that merged after mass transfer, as this would require fine-tuning of the system initial parameters. 
	The initial separation has to be just long enough that the primary reaches the AGB phase, during which some material 
	is transferred onto the companion, before a common-envelope phase shrinks the orbit so much that the system 
	merges within a Hubble time. Furthermore, the merger product would itself end up on the AGB, 
	and the same counter-argument as given above against self-pollution applies.
	Hence, the most likely hypothesis is that these stars were born with enhanced 
	abundances of carbon and $s$-elements from an already enriched interstellar medium. 

	\citet[][]{Choplin2017-2} present a model in which the winds of rapidly-rotating 
	massive stars (also called ``spinstars'') are the cause of this pollution. They
	were able to reproduce the abundances of most detected elements in the four 
	apparently-single CEMP-$s$ stars with a model of a $25\Msun$ star rotating at $70\%$ of the break-up velocity
	at $[\Fe/\Hy]\approx-1.8$ (their Fig.~3). The comparison between the best fits of 
	\cite{Choplin2017-2} and those of \cite{Abate2015-2}, who in their model set A used essentially 
	the same input physics as in our model set M2, indicates that for
	two of these stars (HE~$0206$--$1916$ and CS~$30301$--$015$) the binary mass-transfer model gives
	a better fit to the observed abundances. In addition, in one case (star HE~$1045$+$0225$) both models fail to
	reproduce at the same time the abundances of light elements (carbon, sodium, magnesium) and light $s$-elements 
	(strontium and yttrium), with the spinstar model correctly fitting the $s$-elements and the binary model
	better reproducing the light elements. 
	
	Star CS~$30301$--$015$ is particularly interesting as it has the
	largest set of measured abundances, including heavy $s$-elements up to lead, that can put tighter 
	constrains on the nucleosynthesis models. While the analysis of \cite{HansenTT2016-2} shows no 
	radial-velocity variations within $0.3~\kms$ over a $2,\!300$-day time span, 
	the combination of relatively low abundances of light $s$-elements ($[\ls/\Fe]\approx 0.4$),
	mild enrichment in heavy $s$-elements ($[\hs/\Fe]\approx 1$, though with some scatter)
	and strong lead enhancement ($[\Pb/\Fe]=2$, \citealp{Aoki2002-5}) cannot be reconciled with
	the predictions of the spinstar model and is much more consistent with AGB nucleosynthesis.
	Also, the binary model of \cite{Abate2015-2} for this star predicts an orbital period which
	could be as long as $10^6$~days, which would likely go undetected.
	
	In conclusion, although the spinstar model naturally explains why no orbital motion is detected,
	binary-star models \cite[][and our model set M2]{Abate2015-2} generally predict wide orbits for
	these systems ($P>15,\!000$), which would be difficult to detect. The abundances of more elements
	are necessary to discriminate between a spinstar or AGB origin of the chemical enrichment
	in these stars. In particular, rotating massive stars generally yield more oxygen and sodium compared to AGB stars
	for a given amount of carbon. Also, they produce more light $s$-elements
	(e.g. strontium) than heavy $s$-elements (e.g. barium) and not much lead
	\cite[][]{Frischknecht2012, Choplin2017-2}. By contrast, in AGB stars heavy $s$-elements and lead are usually more
	strongly enhanced than light $s$-elements. When a set of observed abundances of all these elements will 
	be available for these stars, it will be possible to put stronger constraints on their origin.

\section{Summary and conclusions}
\label{concl}

	The motivation for this work is that in population-synthesis models of metal-poor binary systems
	the majority of CEMP-$s$ stars have periods several times longer than in the observed sample of 
	CEMP-$s$ stars compiled by \cite{HansenTT2016-2}. This sample is taken as a reference because
	arguably it is not biased towards a particular period range. In previous studies it has normally
	been assumed that Roche-lobe overflow is unstable when the donor is an AGB star (except in rare
	circumstances), that wind ejection occurs in spherical symmetry, and that the initial distribution
	of orbital separations is flat in the logarithm. This results in about $45\%$ of the synthetic
	CEMP-$s$ stars having periods exceeding $10,\!000$ days, which is currently the longest period
	measured for an observed system. If these wide systems are excluded from the synthetic
	population, the models underestimate the observed CEMP fraction by almost a factor of two.

	In this study we consider several modifications of these standard assumptions and investigate their
	effect on the CEMP period distribution. We show that the stability criterion of Roche-lobe overflow
	plays a role in determining the proportion of CEMP stars that are formed at periods between a few 
	hundred and a few thousand days. However, even if we assume that Roche-lobe overflow from AGB donors 
	is always stable, the consequences for the period distribution of the synthetic CEMP population are
	small, because only a relatively small fraction of all our simulated CEMP stars experience a phase
	of Roche-lobe overflow, between about $15\%$ and $30\%$ in the most extreme case, while the remainder
	form by wind accretion. Hence, to reproduce the observed fraction of CEMP stars, wind mass transfer
	from AGB donors in binary systems has to be efficient. 

	A large uncertainty in our study is the original binary fraction per decade of orbital period in the
	very metal-poor stellar population of the Halo. Assuming that it was similar to that observed at
	higher metallicity in the solar neighbourhood, the constraint placed by the observed period 
	distribution of CEMP-$s$ stars requires that the wind ejected by binary systems has to carry away
	a large amount of angular momentum, up to about ten times higher than in the simplistic case 
	of isotropic wind ejection. At present, state-of-the-art hydrodynamical simulations
	do not predict the high angular-momentum loss necessary to reconcile the results of our 
	population-synthesis models with the observations. However, another possibility is that at very low
	metallicity binary systems formed at periods distributed differently from today. Our simulations show that
	if binary systems are initially formed in a significantly narrower range of periods, up to
	around ten thousand days, then the period distribution of observed CEMP-$s$ stars can be reproduced.

	The CEMP-$s$ star sample of \cite{HansenTT2016-2} contains four stars that appear to be single,
	that is, no evidence for binary-induced radial-velocity variations was found. Our simulations
	show that at least some, and perhaps all, of these could be binaries at periods (much) longer
	than $10,\!000$ days that are observed at unfavourable orbital phases and/or inclinations.
	However, it is hard to reconcile at the same time both the substantial number of apparently
	single stars and the fact that only one detected binary has a (still unmeasured) period
	exceeding $10,\!000$ days with any of our simulations.

	In conclusion, a combination of significant wind accretion efficiency, higher than the 
	predictions of the canonical Bondi-Hoyle-Lyttleton approximation, strong angular-momentum
	loss carried away by the wind material that escapes the binary system, and possibly an
	initial distribution of orbital separations significantly different from that observed
	in solar-vicinity stars, is required to reproduce the orbital periods of the observed
	population of CEMP-$s$ stars.

\begin{acknowledgements}
We thank the anonymous referee for valuable comments that have helped improve our paper.
CA is the recipient of an Alexander von Humboldt Fellowship. RJS is the recipient of a Sofja Kovalevskaja Award from the Alexander von Humboldt Foundation. 
\end{acknowledgements}


\Online
%
\appendix

\section{Fit to ballistic calculations}
\label{appendixA}

\cite{Brookshaw1993} compute the average specific angular momentum $\langle j_w \rangle$ of test particles ejected from the surface of a star and subsequently lost from a binary system, after following their ballistic trajectories in the binary potential. They present their results in terms of a quantity
\begin{equation} \label{eq:app1}
h_\mathrm{cm} = (1 + Q) \frac{\langle j_w \rangle}{a^2 \Oorb}~,
\end{equation}
which is related to our parameter $\eta$ as $h_\mathrm{cm} = \eta \, (1+Q)$. However, in their calculations the mass-losing star co-rotates with the orbit, so that $h_\mathrm{cm}$ includes both orbital and spin angular-momentum loss. Since the $\texttt{binary\_c}$ code already accounts for spin angular-momentum loss and spin-orbit coupling explicitly, we should avoid including this effect twice. Thus $\eta$ should include only the orbital angular-momentum loss. We expect the spin angular momentum loss to contribute a term equal to $\langle j_\mathrm{rot} \rangle = \frac{2}{3} R^2 \Oorb$ for a co-rotating star of radius $R$, which implies that
\begin{equation} \label{eq:app2}
\eta = \frac{h_\mathrm{cm}}{1+Q} - \frac{2}{3} \frac{R^2}{a^2}~. 
\end{equation}
Table~13 of \cite{Brookshaw1993} presents values of $h_\mathrm{cm}$ for various mass ratios and Roche-lobe filling factors, $\Psi = R/R_\mathrm{L}$, of the mass-losing star. Because in these calculations the particles are ejected at very high velocity, we expect the Jeans approximation to be valid, i.e. we expect $\eta$ to be equal to $\eiso = (1+Q)^{-2}$ (Eq.~\ref{eq:jeans}). We verified that this is indeed the case when applying Eq.~(\ref{eq:app2}) to all the values in Table~13.

\begin{figure}
\includegraphics[width=0.5\textwidth]{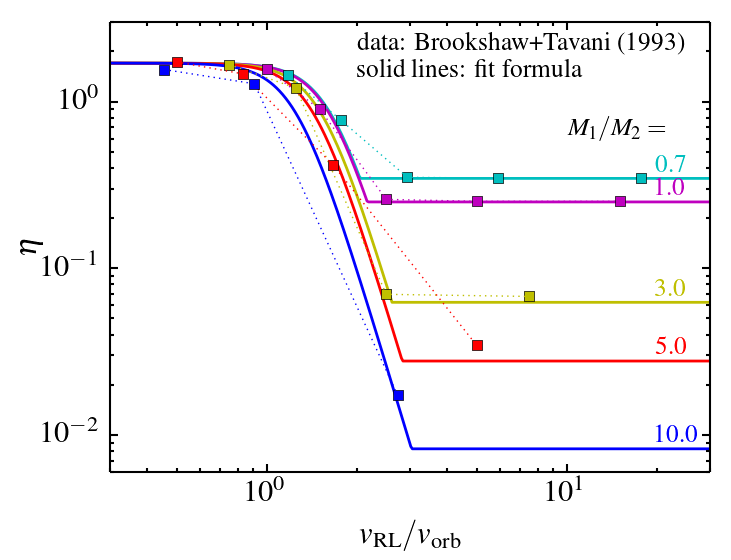}
\caption[]{Results from Table~1 of \cite{Brookshaw1993} for several mass ratios $Q$ (coloured squares) as a function of $v_\mathrm{in}/v_\mathrm{orb}$. Solid lines show the fitting formula Eq.~(\ref{eq:BT93}) for the corresponding mass ratios.}
\label{fig:BT93}
\end{figure}

We thus proceed to use Eq.~(\ref{eq:app2}) to study the dependence of $\eta$ on the particle injection velocity $v_\mathrm{in}$. We use the results of Table~1, in which isotropic mass loss from the surface of a Roche-lobe filling star is assumed. Applying Eq.~(\ref{eq:app2}) in this case is not quite correct because it neglects the non-spherical shape of a Roche-lobe filling star, but it is accurate enough for our purposes. Table~1 tabulates $h_\mathrm{cm}$ as a function of the mass ratio $Q$ and the parameter $V = v_\mathrm{in}/v_\mathrm{orb,d}$, where $v_\mathrm{orb,d} = \vorb/(1+Q)$ is the orbital velocity of the donor star around the centre of mass and $\vorb$ is the relative orbital velocity of the binary. Fig.~\ref{fig:BT93} shows the corresponding $\eta$ as a function of $v_\mathrm{in}/v_\mathrm{orb}$ for several values of $Q$, together with the fitting formula of Eq.~(\ref{eq:BT93}). In using the latter we have identified $\vw$ with $v_\mathrm{in}$. For $v_\mathrm{in} \ga 3\vorb$ the results of the ballistic calculations reproduce the Jeans mode, $\eiso$ (horizontal parts of the curves) while for $v_\mathrm{in} \la \vorb$ the results converge to a constant value of $\eta \approx 1.7$, independent of mass ratio.

\section{Reduced time of radial-velocity monitoring}
\label{appendixB}

\begin{figure}[ht]
		\vspace{1cm}
	    \includegraphics[width=0.5\textwidth]{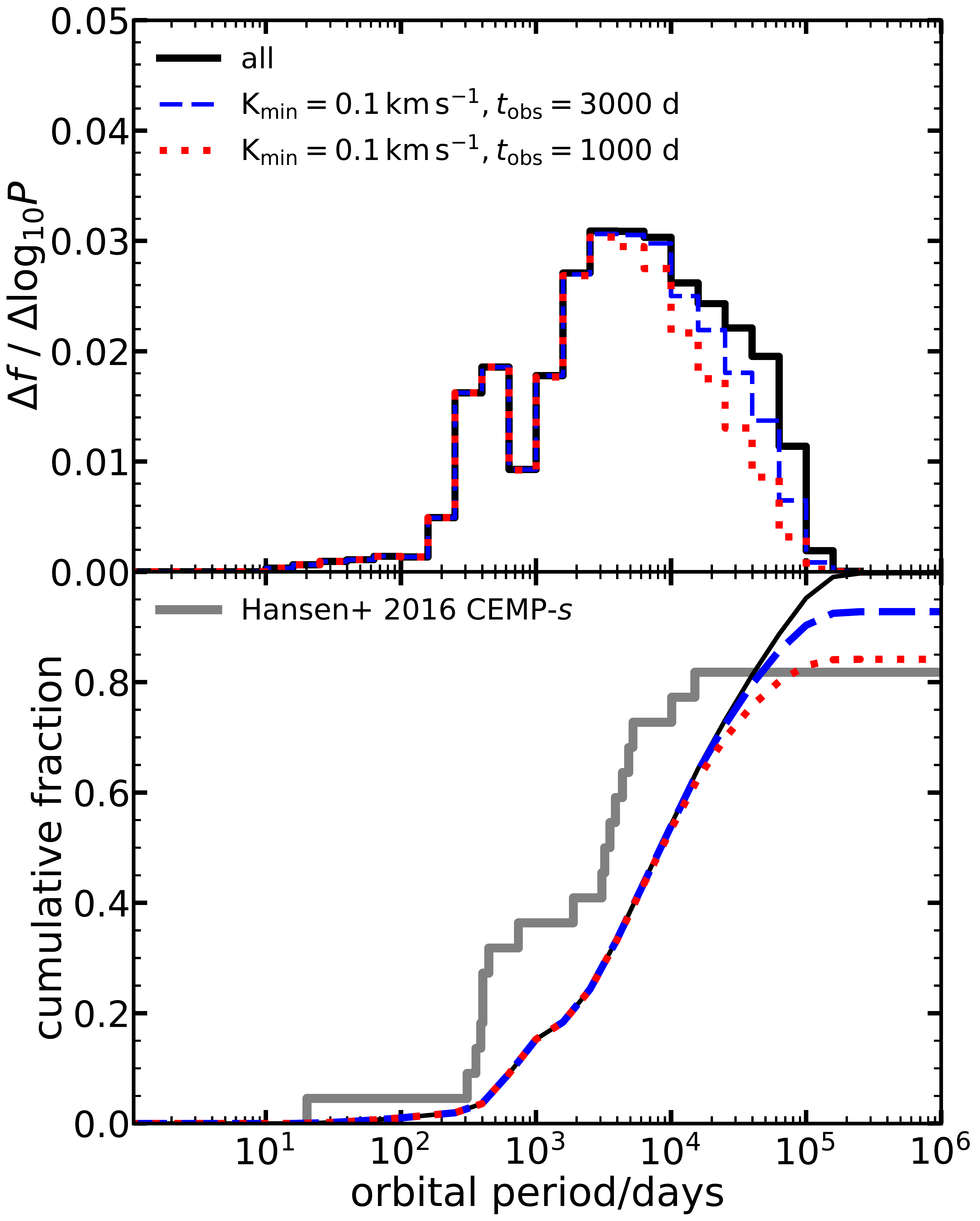}
	    \caption{As Fig. \ref{fig:qcrit} for models with different time-span of the radial-velocity monitoring.
	    		The solid-black line shows the period distribution of all the CEMP-$s$ stars in our simulation. %
	    		The blue-dashed and red-dotted lines are computed with a detection threshold $\Kmin=0.1\,\kms$
	    		and a time-span of $3,\!000$ and $1,\!000$ days, respectively.
	    		}
	\label{fig:baseline}
	\end{figure}

	Figure \ref{fig:baseline} illustrates how the assumption of two different radial-velocity monitoring times, 
	namely $t_{\mathrm{obs}}=3,\!000$ and $t_{\mathrm{obs}}=1,\!000$ days (blue-dashed and red-dotted lines, respectively) 
	modifies the period distribution of the synthetic CEMP-$s$ systems. The period distributions are computed
	with our default model set M2 and a detection threshold of $\Kmin=0.1\,\kms$. The period distribution of
	all our synthetic CEMP-$s$ stars is also shown for comparison (black-dashed line). %

\end{document}